\documentclass[10pt,journal,twocolumn]{IEEEtran}
\usepackage{amssymb}
\usepackage{graphicx}
\usepackage{amsmath}
\usepackage{epsf}
\usepackage{mathrsfs}
\usepackage{cite}
\usepackage{dsfont}
\usepackage{float}
\usepackage{color}

 \usepackage[usenames,dvipsnames]{pstricks}
 \usepackage{epsfig}
 \usepackage{pst-grad} 
 \usepackage{pst-plot} 
 \usepackage{pst-node}
 \usepackage{pst-3d}
 \usepackage{pstricks-add}

\newtheorem{theorem}{Theorem}

\newtheorem{definition}{Definition}

\newtheorem{example}{Example}

\newtheorem{remark}{Remark}

\def\antenna{%
\begin{pspicture}(1,1)
\pstriangle[gangle=-180.0](0,0.25)(0.6,0.25)
\psline(0,0)(0,-0.60)
\psline(-0.22,-0.60)(0.22,-0.60)
\end{pspicture}
}

\def\mydef#1#2{
\begin{figure}[H]
\centering
\begin{tabular}{l}
  \parbox{7cm}{\bf{#1}}\\
  \hline
  \parbox{7cm}{\vspace{.15cm}\emph{#2}}
\end{tabular}
\end{figure}
}

\begin{document}
\title{Real Interference Alignment: Exploiting the Potential of Single Antenna Systems}


\author{ Abolfazl~Seyed~Motahari$^{\dagger}$, ~Shahab Oveis-Gharan$^{\dagger}$,\\~Mohammad-Ali~Maddah-Ali$^{\dagger \dagger}$,~and~Amir~Keyvan~Khandani$^{\dagger}$
\\\small $^{\dagger}$  Department of Electrical and Computer Engineering,  University of Waterloo
\\\small Waterloo, ON, Canada N2L3G1
\\ \{abolfazl,shahab,khandani\}@cst.uwaterloo.ca
\\ \small $^{\dagger\dagger}$ Department of Electrical Engineering and Computer Sciences, University of California-Berkeley
\\ \small Berkeley, CA, USA
\\ maddah-a@eecs.berkeley.edu
}



\maketitle

\footnotetext{Financial support provided by Nortel and the
corresponding matching funds by the Natural Sciences and Engineering
Research Council of Canada (NSERC), and Ontario Ministry of Research
\& Innovation (ORF-RE) are gratefully acknowledged.}

\begin{abstract}
In this paper, the available spatial Degrees-Of-Freedoms (DOF) in
single antenna systems is exploited. A new coding scheme is proposed
in which several data streams having fractional multiplexing gains
are sent by transmitters and interfering streams are aligned at
receivers. Viewed as a field over rational numbers, a received
signal has infinite fractional DOFs, allowing simultaneous
interference alignment of any finite number of signals at any finite
number of receivers. The coding scheme is backed up by a recent
result in the field of Diophantine approximation, which states that
the convergence part of the Khintchine-Groshev theorem holds for
points on non-degenerate manifolds. The proposed coding scheme is
proved to be optimal for three communication channels, namely the
Gaussian Interference Channel (GIC), the uplink channel in cellular
systems, and the $X$ channel. It is proved that the total DOF of the
$K$-user GIC is $\frac{K}{2}$ almost surely, i.e. each user enjoys
half of its maximum DOF. Having $K$ cells and $M$ users within each
cell in a cellular system, the total DOF of the uplink channel is
proved to be $\frac{KM}{M+1}$. Finally, the total DOF of the $X$
channel with $K$ transmitters and $M$ receivers is shown to be
$\frac{KM}{K+M-1}$.

\end{abstract}


\section{Introduction}
\PARstart{T}{ime}, frequency, and space are natural resources in
wireless systems. While time and frequency are two global resources
independent of systems' topologies, space is a local resource
related to the number of antennas incorporated in transceivers.
Spectrum sharing is known as a key solution to time/frequency
allocation among several users. To avoid interference in the system,
orthogonal schemes do not allow different transmissions overlap in
time or frequency. Orthogonal schemes fall short of achieving high
throughput in dense networks because allowing for multi-user
interference is proved to be optimal in such networks.

Achieving the optimum throughput of a system requires efficient
interference management. Interference alignment is a type of
interference management that exploits spatial Degrees-Of-Freedoms
(DOF) available at transmitters and receivers. Interference
alignment makes the interference less damaging by merging the
communication dimensions occupied by interfering signals. In
\cite{maddahali2008com}, Maddah-Ali, Motahari, and Khandani
introduced the concept of interference alignment and showed its
capability in achieving the full Degrees-Of-Freedom (DOF) for
certain classes of two-user $X$ channels. Being simple and at the
same time powerful, interference alignment provided the spur for
further research. Besides lowering the harmful effect of the
interference, interference alignment can be applied to provide
security in networks, c.f. \cite{Ozan-gamal-lai-poor}.

The study of interaction between two users sharing the same channel
goes back to Shannon's work on the two-way channel in
\cite{TWOWAY:SHANNON}. His work was followed by several researchers
and the two-user interference channel emerged as the fundamental
building block in dealing with interference in networks.

Although partial capacity results on the interference channel are
recently derived, c.f. \cite{abolfazl,Shang,Annapureddy}, the
problem of characterizing the capacity region of the Gaussian
Interference Channel (GIC) is still open. In~\cite{Etkin},  it is
shown that in the two-user GIC, the Han-Kobayashi (HK) scheme
\cite{Han-kobayashi} achieves within one bit of the capacity region,
as long as the interference from the private message in the HK
scheme is designed to be below the noise level.

It turns out that moving from the two-user scenario to a larger
number of users is a challenging task. Indeed, for $K$-user GIC
($K>2$), the Han-Kobayashi approach of interference management is
not enough and we need to incorporate the interference alignment in
the signaling.

Interference alignment in $n$-dimensional Euclidean spaces for
$n\geq 2$ is studied by several researchers, c.f.
\cite{maddahali2008com,jafar2008dfr,cadambe2008iaa,cadambe2008dfw}.
In this method, at each receiver a subspace is dedicated to
interference, then the signaling is designed such that all the
interfering signals are squeezed in the interference sub-space. Such
an approach saves some dimensions for communicating desired signal,
while keeping it completely free from the interference. Using this
method, Cadambe and Jafar showed that, contrary to the popular
belief, a $K$-user Gaussian interference channel with varying
channel gains can achieve its total DOF, which is $\frac{K}{2}$.
Later, in \cite{Bobak}, it is shown that the same result can be
achieved using a simple approach based on a particular pairing of
the channel matrices. The assumption of varying channel gains,
particularly noting that all the gains should be known at the
transmitters, is unrealistic, which limits the application of these
important theoretical results in practice. This paper aims to remove
this shortcoming.

In~\cite{Bresler-Parekh-tse}, followed up by
\cite{Sridharan,sridharan3llc}, interference alignment is applied in
single antenna systems. In~\cite{Bresler-Parekh-tse}, it is shown
that lattice codes, rather than random Gaussian codes, are essential
parts of signaling for three-user time-invariant GICs. In
\cite{Sridharan}, after aligning interference using lattice codes,
the aggregated signal is decoded and its effect is subtracted from
the received signal. In fact, \cite{Sridharan} shows that the very
strong interference region of the $K$-user GIC is strictly larger
than the corresponding region when alignment is not applied. In
their scheme, to make the interference less severe, transmitters use
lattice codes to reduce the code-rate of the interference, which
guarantees decodability of the interference at the receiver. In
\cite{sridharan3llc}, Sridharan et al. showed that the DOF of a
class of 3-user GICs with fixed channel gains can be greater than 1.
This result was obtained using layered lattice codes along with
successive decoding at the receiver.

In  \cite{Etkin-Ordentlich} and \cite{abolfazl-shahab-amir}, the
results from the field of Diophantine approximation in Number Theory
are used to show that interference can be aligned using properties
of rational and irrational numbers and their relations. They showed
that the total DOF of some classes of time-invariant single antenna
interference channels can be achieved. In particular, Etkin and
Ordentlich in \cite{Etkin-Ordentlich} proposed an upper bound on the
total DOF, which accounts for the properties of channel gains with
respect to being rational or irrational.  Using this upper bound,
surprisingly, they proved that the DOF is everywhere discontinuous
for the class of channels under investigation.

The channels considered in \cite{Etkin-Ordentlich} and
\cite{abolfazl-shahab-amir} are special in the sense that signals
not intended for a given receiver are aligned by the channel.
Therefore, signaling design is not required due to the nature of the
channel. The first example of interference alignment in
one-dimensional spaces, which requires signaling design, is
presented in \cite{abolfazl-real}. Using irrational numbers as
transmit directions and applying Khintchine-Groshev theorem,
\cite{abolfazl-real} shows the two-user $X$ channel achieves its
total DOF. This is the first channel in which no variations in
coefficients over time or frequency and no multiple antennas are
required to achieve the total DOF. This is because rational
dimensions in one-dimensional spaces can play the role of real
dimensions in more-than-two dimensional spaces. In this paper, we
take one step forward and prove that the total DOF of the $K$-user
GIC, the uplink channel in cellular systems, and the $X$ channel can
be achieved without the need for channel variation over
time/frequency/space.

This paper is organized as follows: in Section \ref{sec main
theorem}, the main theorem of this paper is stated and some
discussions follow. In Section \ref{sec main ideas}, the main ideas
incorporated in the proposed coding scheme are presented. Moreover,
several examples are provided to shed light on the ideas. In Section
\ref{sec groshev}, some background on the field of Diophantine
approximation and, in particular, Khintchine-Groshev type theorems
are presented. Section \ref{sec coding1} describes the coding scheme
used to prove the main theorem. Moreover, the performance analysis,
based on recent results in the field of Diophantine approximation,
is presented. In Section \ref{sec k-user}, the total DOF of the
$K$-user GIC is derived. In Section \ref{sec uplink}, it is proved
that the uplink channel in cellular systems has $\frac{KM}{M+1}$
DOF, where $K$ is the number of cells and $M$ is the number of users
within each cell. In Section \ref{sec X channel}, the total DOF of
the $K\times M$ $X$ channel is derived. Finally, Section \ref{sec
conclusion} concludes the paper.

\textbf{Notation}: $\mathbb{R}$, $\mathbb{Q}$, $\mathbb{N}$
represent the set of real, rational, and nonnegative integers,
respectively. For a random variable $X$, $E[X]$ denotes the
expectation value. $(a,b)_{\mathbb{Z}}$ denotes the set of integers
between $a$ and $b$.

\section{Main Contributions and Discussions}\label{sec main theorem}

\subsection{Main Results}
In this paper, the total DOFs of three channels, namely the $K$-user
GIC, the uplink channel in cellular systems, and the $K\times M$ $X$
channel, are characterized using a new coding scheme.
\begin{theorem}\label{main}
The total DOF of the $K$-user GIC with real and time invariant channel coefficients is $\frac{K}{2}$ for almost all channel realizations.
\end{theorem}

\begin{theorem}\label{main 2}
The total DOF of a cellular system consisting of $K$ cells and $M$ users within each cell is $\frac{KM}{M+1}$ for almost all channel realizations.
\end{theorem}

\begin{theorem}\label{main 3}
The total DOF of the $K\times M$ $X$ channel with real and time invariant channel coefficients is $\frac{KM}{K+M-1}$ for almost all channel realizations.
\end{theorem}

\subsection{Real Interference Alignment}
The available DOF of the systems having multiple-antenna,
time-varying, and/or frequency-selective channels can be efficiently
exploited by choosing appropriate signaling directions to maximize
the channel gains and avoiding or aligning interference. We refer to
the alignment scheme incorporating directional signaling as
\emph{vector alignment}.  In contrary, it was commonly believed that
time-invariant frequency-flat single-antenna channels are
restrictive in the sense that they prevent us to incorporate vector
alignment. Here, we develop a machinery that transforms the
single-antenna systems into pseudo multiple-antenna systems with
infinite-many pseudo antennas. Indeed the number of available
dimensions in the resultant pseudo multiple-antenna systems is,
roughly speaking, as many as rationally-independent irrational
numbers. We see that the pseudo multiple-antenna channels mimics the
behavior of real multi-dimensional systems (in time/frequency/space)
and, for example, allows us to simultaneously align interference at
all receivers of static single-antenna channels. We refer to the
alignment scheme applicable in single antenna systems as \emph{real
alignment}.

\subsection{Almost All vs All Cases}
In the statement of the theorem, it is emphasized that the total
DOFs of the $K$-user GIC, the uplink channel in a cellular system,
and the $X$ channel are achievable for almost all channel
realizations. It means the collection of all possible channel
realizations in which the total DOF may not be achieved has measure
zero. In other words, if all channel gains are drawn independently
from a random distribution then almost surely the channel has the
desired properties  required for achieving the total DOF.

In the case of the $K$-user GIC if all channel gains are rational,
then the total DOF is strictly less than $\frac{k}{2}$. This is due
to the recent upper bound on the total DOF obtained by Etkin and
Ordentlich in \cite{Etkin-Ordentlich}. This result, together with
Theorem \ref{main}, implies that the total DOF of the channel is
everywhere discontinuous with respect to channel coefficients. This
is due to the fact that for any set of channel gains one can find a
set of rational numbers arbitrarily close to it.  This behavior is
unique to this channel (or related networks with single antennas).
In fact, almost all of the total DOFs obtained for Multiple Input
Multiple Output (MIMO) systems are discontinuous at a point or on a
set of measure zero. However, none of them are everywhere
discontinuous.


Other than rational channel gains, infinitely many channel
realizations are not covered by the theorems. However, it cannot be
concluded that for these realizations the total DOFs are not
achievable. In fact, it is proved that there are some cases where
the total DOFs can be achieved and those cases are out of the scope
of the theorems, c.f.,
\cite{Etkin-Ordentlich,abolfazl-shahab-amir,abolfazl-real}. As an
example, the total DOF of the $K$-user GIC can be achieved by using
a single layer constellation  at transmitters in the special case
where all cross gains are rational numbers and all direct gains are
algebraic irrationals (this is the case for almost all
irrationals)\cite{Etkin-Ordentlich}. This is due to the fact that
cross gains lie on a single rational dimension and therefore, the
effect of the interference caused by several transmitters behaves as
that of interference caused by a single transmitter. Using a single
data stream, one can deduce that the multiplexing gain of
$\frac{1}{2}$ is achievable for each user.


\subsection{Time Varying versus Time-Invariant Channels}
Cadambe and Jafar in their papers \cite{cadambe2008iaa} and
\cite{cadambe2008dfw} proved that the total DOFs of the time-varying
$K$-user GIC and $X$ channel can be achieved. They showed  that the
variation of the channel in time, if it is fast enough to be assumed
independent, provides enough freedom to align the interference.
However, such an assumption about the variation of wireless channels
is not practically realistic. Moreover, it imposes an inadmissible
delay on the system, noting that wireless channels are changing
slowly.



Here, we propose a signaling scheme that achieves the total DOFs in
almost all realizations of the channel without imposing any delay to
the system or requiring channel variation. Indeed, the channel can
be static over time and still it is possible to achieve the total
DOFs of the channels.




%


\subsection{MIMO and Complex Coefficients Cases}
Let us consider the $K$-user MIMO GIC where each node in the network
is equipped with $M$ antennas. The upper bound on the total DOF
states that at most $\frac{MK}{2}$ is achievable for this channel.
Except for the three-user case where Cadambe and Jafar in
\cite{cadambe2008iaa}, through explicit interference alignment,
showed that $\frac{3M}{2}$ is achievable, the total DOF of $K$-user
MIMO GIC with static channel states is not considered in the
literature. Again, if we assume time-variant channels, however, this
upper bound can be achieved, see \cite{cadambe2008iaa}.


The applicability of Theorem  \ref{main} is not restricted to the
single antenna case. In fact, we can also show that for the $K$-user
MIMO GIC the total DOF of the channel can be achieved for almost all
cases. This can be proved by simply viewing a single user as $M$
virtual users in which a transmit antenna is paired with a receive
antenna. Using separate encoding (resp. decoding) at all transmit
(resp. receive) antennas, the channel becomes a $MK$-user single
antenna GIC. Applying the theorem to this channel, we conclude that
the total of $\frac{MK}{2}$ is achievable and this meets the upper
bound. In \cite{Akbar-Abolfazl-Amir}), the total DOF of the $K$-user
IC is obtained for the case where the numbers of transmit and
receive antennas are different.

Needless to say, Theorem \ref{main} is also applicable to channels
(either single or multiple antennas) with complex coefficients. In
fact, the real and imaginary parts of the input and the output can
be paired. This converts the channel to $2K$ virtual users.
Therefore, the total DOF of the channel can be achieved by a simple
application of the theorem. It is worth noting that joint processing
between all antennas and/or real-imaginary parts at a transmitter
increases the achievable sum rate of the channel. However, at high
SNR regimes this increase vanishes and the total DOF of the channel
can be achieved by separate coding over all available dimensions.

The total DOF of the $X$ channel with complex coefficients follows
similar behavior, but it can not be derived by pairing. In fact, a
simple extension of the coding proposed in this paper results in the
total DOF of this channel \cite{maddahalidof}.

\section{Main Ideas and Basic Examples}\label{sec main ideas}
In this section, we review some important features of the real
interference alignment introduced in \cite{abolfazl-real} and extend
its application to more general cases. To clarify basic ideas, we
rely on some simple examples and provide only rough reasoning for
rationality of the schemes. Unless otherwise stated, the following
assumptions are in place throughout this section. \mydef{Generic
Assumptions}{\begin{enumerate}
\item All channels are additive.
\item The received signals are corrupted by unit variance additive Gaussian noise.
\item All transmitters are subject to power constraint $P$.
\end{enumerate}}

In \cite{abolfazl-real}, constellation points carved from integers
are used to construct a code for transmission of a given data
stream. Carrying multiple data streams, a transmitter designs its
\emph{transmit constellation} based on a linear combination of
constellations designed for individual data streams. Since all
transmitters use a set of finite points as the input symbols, the
received symbol before corruption by additive noise is also a finite
set, which is called the \emph{received constellation}.

It will be shown that the performance of the system is highly
related to the design of transmit constellations. In order to focus
on the important aspects of the optimum constellation design, we
bypass the effect of error correction codes and assume that
receivers can remove the additive noise under the following
condition:

\mydef{Noise Removal}{A receiver can completely remove the noise if
the minimum distance of the received constellation points is greater
than $\sqrt{N}$, where $N$ is the noise variance.}

The preceding assumption is by no means correct. However, it
provides accurate estimates of the total DOFs of the systems under
investigation. In the following sections, we will explain that if
the minimum distance of the received constellation is of order of
$\sqrt{N}P^{\epsilon}$ for any $\epsilon>0$, then a code with rate
arbitrary close to the size of the transmit constellation exists
such that the noise can be completely removed from the received
signal. To see the power of the above assumption, we look at the
following examples.

\def\antenna{\psline(0.05,0)(.35,0)
\psline(.2,0)(.2,0.5)
\pspolygon(.2,0.5)(.4,.7)(0,.7)}

\begin{figure}
\centering
\scalebox{1}
{
\begin{pspicture}(1,0)(6,1)
\rput(2,0){\antenna}
\rput(4,0){\antenna}
\psline{->}(2.5,.4)(3.9,.4)
\psline{->}(1.2,.4)(2,.4)
\rput(1.6,.6){\scriptsize $x$}
\psline{->}(4.4,.4)(5.2,.4)
\rput(5.4,.6){\scriptsize $y=x+z$}

\rput(5,0){
\psline(0,0)(2,0)
\psline(.2,-.1)(.2,.1)
\psline(.6,-.1)(.6,.1)
\psline(1,-.1)(1,.1)
\psline(1.4,-.1)(1.4,.1)
\psline(1.8,-.1)(1.8,.1)
\psline[linewidth=.01]{<->}(.6,-.15)(1,-.15)
\rput(.9,-.3){\scriptsize $d_{\min}$}}

\rput(-0.5,0){
\psline(0,0)(2,0)
\psline(.2,-.1)(.2,.1)
\psline(.6,-.1)(.6,.1)
\psline(1,-.1)(1,.1)
\psline(1.4,-.1)(1.4,.1)
\psline(1.8,-.1)(1.8,.1)
}

\end{pspicture}
}
\caption{A point-to-point communication system. The receive constellation is the same as the transmit constellation.}\label{fig point-to-point}
\end{figure}
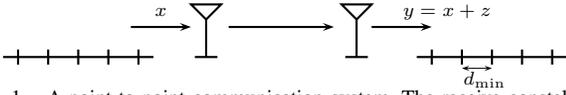

\begin{figure}
\centering
\scalebox{1}
{
\begin{pspicture}(1,1)(5.6,4)
\rput(2,1){\antenna}
\rput(2,2){\antenna}
\rput(2,3){\antenna}
\rput(4.2,2){\antenna}

\psline{->}(1.2,1.4)(2,1.4)
\rput(1.6,1.6){\scriptsize $x_3$}
\psline{->}(1.2,2.4)(2,2.4)
\rput(1.6,2.6){\scriptsize $x_2$}
\psline{->}(1.2,3.4)(2,3.4)
\rput(1.6,3.6){\scriptsize $x_1$}

\psline{->}(4.6,2.4)(5.4,2.4)
\rput(5,2.6){\scriptsize $y$}

\cnode[linecolor=white](2.4,3.4){.1}{T1}
\cnode[linecolor=white](2.4,2.4){.1}{T2}
\cnode[linecolor=white](2.4,1.4){.1}{T3}
\cnode[linecolor=white](4.2,2.4){.1}{R}
\ncline{->}{T1}{R}
\ncput*{\scriptsize 1}
\ncline{->}{T2}{R}
\ncput*{\scriptsize a}
\ncline{->}{T3  }{R}
\ncput*{\scriptsize b}

\end{pspicture}
}
\caption{A multiple access channel.}\label{fig Mac}
\end{figure}
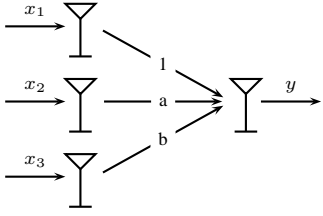

\begin{example}[Point-to-point communication]
A single user channel is shown in Figure \ref{fig point-to-point}.
Given an integer $Q$, the transmit constellation
$\mathcal{U}=(-Q,Q)_{\mathbb{Z}}=\{-Q,-Q+1,\ldots,Q-1,Q\}$ is used
for transmission of a single data stream. Since it is assumed that
the additive noise has unit variance and the minimum distance in the
received constellation, which is the same as the transmit
constellation, is also one, the noise can be removed from the
received signal. Therefore, $R\approx\log(2Q-1)$ is achievable for
the channel. On the other hand, the input power is less than $Q^2$.
Hence, $P=Q^2$. The multiplexing gain associated with the data
stream can be computed as
\begin{equation}
r=\lim_{P\rightarrow\infty}\frac{R}{0.5\log P}=1.
\end{equation}
\end{example}

\begin{example}[Multiple Access Channel]
A multiple access channel with three users is shown in Figure
\ref{fig Mac}. The channel can be modeled as
\begin{equation*}
y=x_1+ax_2+bx_2+z.
\end{equation*}
It is assumed that $a$ and $b$ are two real numbers. Moreover, let
us assume that all three users communicate with the receiver using a
single data stream. The data streams are modulated by the
constellation $\mathcal{U}=A(-Q,Q)_{\mathbb{Z}}$ where $A$ is a
factor controlling the minimum distance of the received
constellation.

The received constellation consists of points representable by
$A(u_1+au_2+bu_3)$ where $u_i$s are integer. Let us choose two
distinct points $v_1=A(u_1+au_2+bu_3)$ and $v_2=A(u'_1+au'_2+bu'_3)$
in the received constellation. The distance between these two points
is $d=A|(u_1-u'_1)+a(u_2-u'_2)+b(u_3-u'_3)|$. Khintchine-Groshev
theorem (see Section \ref{sec groshev}, Theorem \ref{thm khintchine
groshev}) provides us a lower bound on any linear combination of
integers. Using the theorem, one can obtain $d_{\min}\approx
\frac{A}{Q^{2}}$ where $d_{\min}$ is the minimum distance in the
received constellation. To be able to remove the noise,
$d_{\min}=1$. Hence, $A\approx Q^2$. In a noise-free environment,
the receiver can decode  the three messages if there is a one-to-one
map from the received signal to the transmit constellation.
Mathematically, one can satisfy this condition by enforcing the
following:

\mydef{Separability Condition}{The receiver is able to decode the
three messages if $a$ and $b$ are rationally independent. In other
words, $p_1+ap_2+bp_3=0$ has no non-trivial solution in integers
$p_1$, $p_2$, and $p_3$.}

Having the above condition, the receiver can decode all three
messages. To calculate User $i$'s rate $R_i=\log(2Q-1)$ in terms of
$P$, we need to find a relation between $Q$ and $P$. Due to the
power constraints, we have $P=A^2Q^2$. We showed that $A\approx
Q^2$. Therefore, $P\approx Q^6$. Hence, we have
\begin{equation}
r_i=\lim_{P\rightarrow\infty}\frac{R_i}{0.5\log P}=\frac{1}{3}.
\end{equation}
\end{example}

Two facts are hindered in the preceding example. First,
Khintchine-Groshev theorem is not valid for all possible values of
$a$ and $b$. In fact, there are infinitely many cases that are not
addressed in the theorem (see Figure \ref{fig bad-events}). However,
the theorem asserts that the measure of these points is zero. In
other words, for any smooth probability distribution on the pair of
$(a,b)$, the probability that the theorem holds is one. Second, the
separability condition does not hold in general. In fact, this
condition holds again with measure one. Hence, we can conclude:

\mydef{Achievablity for Almost All Cases}{The proofs presented in
this paper are based on the separability condition and
Khintchine-Groshev type theorems. Therefore, all results are valid
for almost all channel realizations.} 

As mentioned in the previous example, the pair $(a,b)$ can possibly
take all vectors in $\mathbb{R}^2$. Let us assume that $a$ and $b$
have a relation. For instance, $b$ is a function of $a$, say
$b=a^2$. In this case, the pair $(a,b)$ lies on a one dimensional
manifold in $\mathbb{R}^2$, see Figure \ref{fig bad-events}. Since
the manifold itself has measure zero, Khintchine-Groshev theorem can
not be applied directly. For such cases, however, there is an
extension to Khintchine-Groshev theorem (see Section \ref{sec
groshev}, Theorem \ref{groshev-type}) which states that the same
lower bound on the minimum distance can be applied when coefficients
lie on a non-degenerate manifold and, in fact, the measure of points
not satisfying the theorem is zero.

\mydef{Non-degenerate Manifolds \cite{groshev-Beresnevich}}{Let
$U\subset\mathbb{R}^d$ be an open set. The function $f:U\rightarrow
\mathbb{R}^n$ is $l$-non-degenerate at $x_0\in U$ if
\begin{enumerate}
\item $f$ is $l$ times continuously differentiable on some sufficiently small ball centered at $x_0$.
\item Partial derivatives of $f$ at $x_0$ of orders up to $l$ span $\mathbb{R}^n$.
\end{enumerate}
The function $f$ is non-degenerate at $x_0$ if it is
$l$-non-degenerate at $x_0$ for some $l\in N$. We say that $f$ is
non-degenerate if it is non-degenerate almost everywhere on $U$.}

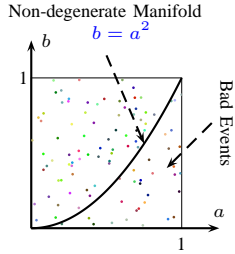
\begin{figure}
\centering
\scalebox{1}
{
\begin{pspicture}(0,0)(3,3)
 \rput(-.1,2){\scriptsize 1}
 \rput(2,-.2){\scriptsize 1}
 \psline{->}(0,2.5)
 \psline{->}(2.5,0)
 \rput(2.5,.2){\scriptsize $a$}
 \rput(.2,2.5){\scriptsize $b$}
 \psRandom[dotsize=1pt,color,fillstyle=solid,randomPoints=100](0,0)(2,2){\psframe[linewidth=0](0,0)(2,2)}
 \psline[linestyle=dashed]{->}(2.4,1.4)(1.8,.8)
 \rput{-90}(2.6,1.4){\scriptsize Bad Events}
 \rput(1.2,2.75){\parbox{3cm}{\scriptsize  Non-degenerate Manifold \\ \centerline{$\textcolor[rgb]{0.00,0.00,1.00}{b=a^2}$}}}
 \psline[linestyle=dashed]{->}(1,2.3)(1.5,1.125)
 \psdot[dotsize=1pt](1.4,1.3)
 \psplot[plotpoints=100]{0}{2}{x x mul 2 div}
\end{pspicture}
}
\caption{There are infinitely many points on the $a-b$ plane with
measure zero that are not addressed by the Khintchine-Groshev
theorem (these are called bad events). The curve $b=a^2$ is a
non-degenerate manifold and by the extension of Khintchine-Groshev
theorem, the measure of bad events is zero on the curve. }\label{fig
bad-events}
\end{figure}

The preceding example can be extended to the multi-user multiple
access channel. The following statement presents the result:

\mydef{Achievable DOF in a Multiple Access Channel}{In a multiple
access channel with $K$ users, each user enjoys $\frac{1}{K}$ of the
total DOF provided that the channel coefficients are independent
over rational numbers and lie on a non-degenerate manifold in
$\mathbb{R}^K$.}

\begin{remark}
Due to the general statement of the achievablity for almost all
cases, the preceding statement about the total DOF of the $K$-user
multiple access channel holds for almost all channels. However, the
capacity region of the $K$-user multiple access channel is
completely characterized and it can be shown that under all
circumstances each user can enjoy $\frac{1}{K}$ of the total DOF. We
will show that the above coding scheme is capable of achieving the
total DOF of channels with interference.
\end{remark}

In the following example, we will look at the two-user $X$ channel
that was originally introduced in \cite{maddahali2008com}. This
example adds two important features to the signaling. First,
multiple data streams are transmitted from each transmitter. Second,
interference alignment is required to achieve the total DOF of the
channel.

\begin{example}[Two-user $X$ Channel \cite{abolfazl-real}]
In the two-user $X$ channel, each transmitter has independent
messages to both receivers, see Figure \ref{fig x channel}. Hence,
each transmitter has two data streams and they need to be
transmitted such that they can be separated in their corresponding
receivers. In \cite{abolfazl-real}, the following signaling is
proposed for the channel:
\definecolor{vcolor}{rgb}{0.00,1.00,1.00}
\definecolor{ucolor}{rgb}{1.00,0.50,1.00}
\begin{IEEEeqnarray*}{rl}
x_1&=h_{22} \textcolor{red}{u_1}+h_{12} \textcolor{blue}{v_1},\\
x_2&=h_{21} \textcolor{red}{u_2}+h_{11} \textcolor{blue}{v_2},
\end{IEEEeqnarray*}
where $u_1,u_2$ and $v_1,v_2$ are data streams intended for the
first and second receivers, respectively. All data streams are
transmitted using the constellation
$\mathcal{U}=A(-Q,Q)_{\mathbb{Z}}$, where $Q$ is an integer and $A$
is the factor controlling the minimum distance of the received
constellation.

The direction used to transmit data streams are chosen based on the
channel coefficients. As we will explain, this choice allows us to
align unwanted signals at receivers. In general, we can state the
following:

\mydef{Transmit Directions Matching the Channel}{To transmit
multiple data streams from a transmitter, channel coefficients are
used as generators for the directions.}

Using the above signaling, the received signal can be written as:
\begin{IEEEeqnarray*}{rl}
y_1&=(h_{11}h_{22}) \textcolor{red}{u_1}+(h_{21}h_{12})\textcolor{red}{u_2} +(h_{11}h_{12})(\textcolor{blue}{v_1}+\textcolor{blue}{v_2})+z_1,\\
y_2&=(h_{21}h_{12}) \textcolor{blue}{v_1}+(h_{11}h_{22})\textcolor{blue}{v_2} +(h_{21}h_{22})(\textcolor{red}{u_1}+\textcolor{red}{u_2})+z_2.
\end{IEEEeqnarray*}
The received signals are linear combinations of three terms in which
two of them are the intended data streams and one is the sum of
interfering signals, see Figure \ref{fig x channel}. Let us focus on
the first receiver. $y_1$ resembles the received signal of a
multiple access channel with three users. However, there is an
important difference between them. In the two-user $X$ channel the
term corresponding to the interfering signals, i.e. $u_3=v_1+v_2$,
is a sum of two data streams. However, we claim that this difference
does not change considerably the minimum distance of the received
constellation, i.e. $d_{\min}$. Recall that Khintchine-Groshev
theorem is used to bound $d_{\min}$. The bound is a function of the
maximum value that the integers can take. The maximum value of $u_3$
is $2AQ$, which is different from a single data stream by a factor
of two. Since this change only affects the constant term of
Khintchine-Groshev theorem, we have $d_{\min}\approx \frac{A}{Q^2}$
and the receiver can decode all data streams if each of them have a
multiplexing gain of $\frac{1}{3}$. Therefore, the multiplexing gain
of $\frac{4}{3}$ is achievable in total, which meets the upper
bound.

\begin{figure}
\centering
\begin{pspicture}(-2.4,-.1)(5,2.8)
\rput(-.3,1.9){\antenna}
\rput(1.8,1.9){\antenna}
\rput(1.8,-0.1){\antenna}
\rput(-.3,-0.1){\antenna}
\rput(-1,2.3){\psline{->}(.7,0) \rput(.3,.2){\scriptsize $x_1$}}
\rput(-1,0.3){\psline{->}(.7,0)\rput(.3,.2){\scriptsize $x_2$}}
\rput{180}(3,2.3){\psline{<-}(.7,0)\rput{180}(.3,-.2){\scriptsize $y_1$}}
\rput{180}(3,.3){\psline{<-}(.7,0)\rput{180}(.3,-.2){\scriptsize $y_2$}}
\cnode(0.2,2.3){0}{T1}
\cnode(0.2,0.3){0}{T2}
\cnode(1.8,2.3){0}{R1}
\cnode(1.8,0.3){0}{R2}
\ncline{->}{T1}{R1}
\ncput*[nrot=:U]{\tiny{$h_{11}$}}
\ncline{->}{T2}{R1}
\ncput*[npos=.7,nrot=:U]{\tiny{$h_{12}$}}
\ncline{->}{T1}{R2}
\ncput*[npos=.7,nrot=:U]{\tiny{$h_{21}$}}
\ncline{->}{T2}{R2}
\ncput*[nrot=:U]{\tiny{$h_{22}$}}

\def\mypoint #1#2#3{\pscircle[linecolor=#3,fillcolor=#3,fillstyle=solid](#1){.2}
\rput(#1){\scriptsize $\textcolor{black}{#2}$}}

\def\mymultipoint #1#2{\rput(#1){
\psframe(-.25,-.25)(.25,1.25)
\mypoint{0,0}{1}{#2}
\mypoint{0,.5}{2}{#2}
}}

\mypoint{-1.4,.3}{v_2}{blue}
\mypoint{-2.1,.3}{u_2}{red}
\mypoint{-1.4,2.3}{v_1}{blue}
\mypoint{-2.1,2.3}{u_1}{red}

\mypoint{4.8,.3}{v_2}{blue}
\mypoint{4.8,2.3}{u_2}{red}
\mypoint{4.1,.3}{v_1}{blue}
\mypoint{4.1,2.3}{u_1}{red}

\rput(4.8,1.8){\scriptsize $\frac{1}{3}$}
\rput(4.1,1.8){\scriptsize $\frac{1}{3}$}
\rput(3.4,1.8){\scriptsize $\frac{1}{3}$}

\rput(4.8,-.2){\scriptsize $\frac{1}{3}$}
\rput(4.1,-.2){\scriptsize $\frac{1}{3}$}
\rput(3.4,-.2){\scriptsize $\frac{1}{3}$}

\pscircle[fillcolor=black,fillstyle=solid](3.4,2.3){.2}
\rput(3.4,2.6){\scriptsize}

\rput(3.4,2.3){
\psframe(-.25,-.25)(.25,.75)
\mypoint{0,0}{v_1}{blue}
\mypoint{0,.5}{v_2}{blue}
}

\rput(3.4,.3){
\psframe(-.25,-.25)(.25,.75)
\mypoint{0,0}{u_1}{red}
\mypoint{0,.5}{u_2}{red}
}

\end{pspicture}
\caption{The two-user $X$ channel. Data streams intended for the
first receiver, $u_1$ and $u_2$, are aligned at the second receiver
occupying one third of the received dimension. Similarly, data
streams intended for the second receiver, $v_1$ and $v_2$, are
aligned at the first receiver occupying one third of the received
dimension. }\label{fig x channel}
\end{figure}
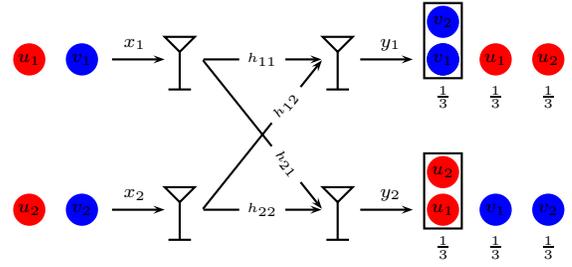
\end{example}

\mydef{Interference Alignment}{Two data streams are aligned at a
receiver if they arrive at the receiver with the same received
direction (coefficient).}

It is interesting to see what Khintchine-Groshev theorem offers when
$v_1$ and $v_2$ receive in different directions. In this case, the
received constellation consists of points representable by a linear
combination of four integers. Therefore, Khintchine-Groshev theorem
gives us $d_{\min}\approx \frac{A}{Q^3}$. Hence, each data stream
can carry information with a multiplexing gain of $\frac{1}{4}$, and
in total, the DOF of 1 is achievable. This means interference
alignment reduces the power of $Q$ in the expression of $d_{\min}$,
which in turn allows achieving higher DOFs.

The signaling proposed for the two-user $X$ channel can be
interpreted as follows. The received signal at each receiver is a
real number, which is a one-dimensional component. One can embed
three rational dimensions, each of which has dimension $\frac{1}{3}$
in this one dimensional space, see Figure \ref{fig x channel}. One
of these dimensions is associated with interference and the other
two with intended signals. Therefore, $\frac{4}{3}$ out of two
dimensions available at both receivers are used for data, which in
turn gives us the total DOF of the channel. In general, we can
state:

\mydef{Rational Dimension Occupation}{If a receiver observes $K$
data stream in $K$ different directions, then each data stream
occupies $\frac{1}{K}$ of the receiver's dimension. Moreover, if
multiple data streams are aligned at a receiver then the dimension
that they occupy is the same as that of a single data streams.}

As above example reveals, available dimensions at all receivers,
like time and frequency, are natural resources in wireless systems.
Interference alignment at all receivers is a way of exploiting the
full potential of these resources by reducing the unused dimensions
at all receivers. In the two user $X$ channel, we have observed that
interfering signals from two different sources can be easily aligned
at a single receiver. Moreover, two interfering streams are received
with the same direction occupying only $\frac{1}{3}$ of the
available dimensions of the receivers. This is in fact the best
efficiency that one can hope for in reducing the number of waste
dimensions. This idea inspires us to define the alignment efficiency
as follows.

\mydef{Alignment Efficiency}{Let us consider that all transmitters
transmit the same number of data streams, say $L_t$, using $L_t$
different directions. Moreover, the maximum dimension occupied by
interference at all receivers is caused by $L_r$ received
directions. The alignment efficiency $\eta$ is defined as the ratio
of $L_t$ and $L_r$ as $$\eta=\frac{L_t}{L_r}.$$ Alignment is called
perfect if $\eta=1$.}

In the two user $X$ channel, we were able to perform perfect
alignment in the system. As the following example shows, however,
this is not the case in general.

\begin{example}[Alignment at two receivers]
Let us consider a communication scenario in which three transmitters
try to align their signals at two different receivers. The channel
is depicted in Figure \ref{fig alignment at two}. In order to shed
light on the alignment part of the signaling, the intended receivers
are removed from the picture.

Alignment can be done at the first receiver by sending a single data
stream with direction $1$ from each of the transmitters; whereas
alignment at the second receiver requires $bc$, $ac$, and $ab$ as
chosen transmit directions for first, second, and third
transmitters, respectively. In general, it is not possible to
simultaneously align three single data streams at two different
receivers. Therefore perfect alignment is not feasible by
transmitting single data streams from each transmitter.

The solution to this problem is partial alignment, which is first
introduced in \cite{cadambe2008iaa}. In this technique, instead of
sending just one data stream, several data streams are transmitted
from each transmitter. The idea is to choose the transmit directions
based on channel coefficients in such a way that the number of
received directions is minimum. For sake of simplicity, we choose
the same directions at all transmitters. Let $\mathcal{T}$ denote
the set of transmit directions. A direction $T\in\mathcal{T}$ is
chosen as a transmit direction if it can be represented as
\begin{equation}
T=a^{s_{1}}b^{s_{2}}c^{s_{3}},
\end{equation}
where $0\leq s_{i}\leq n-1$ for all $i\in\{1,2,3\}$. In this way,
the total number of transmit directions is $L_1=n^3$.

\mydef{Generating Transmit Directions}{Let
$G=\{g_1,g_2,\ldots,g_m\}$ denote a finite set of real numbers. The
set of transmit directions  $\mathcal{G}$  generated by $G$ is the
collection of all real numbers representable by
$$g_1^{s_1}g_2^{s_2}\cdots g_m^{s_m},$$ where $s_i\in \mathbb{N}$
for all $i\in\{1,2,\ldots,m\}$. $\mathcal{G}$ is closed under
multiplication.}

To compute the efficiency of the alignment, one needs to find the
set of received directions in the first and second receivers, which
are denoted by $\mathcal{T}_1$ and $\mathcal{T}_2$, respectively.
Since all transmit directions arrive at the first receiver intact,
$\mathcal{T}_1=\mathcal{T}$.

To compute the set of received directions at the second receiver, we
look at the received directions due to the first, second, and third
transmitters separately. Since all of them are multiplied by $a$,
the received directions due to the first transmitter are of the form
$a^{s_{l}+1}b^{s_{2}}c^{s_{3}}$, where $0\leq s_{i}\leq n-1$ for all
$i\in\{1,2,3\}$. Similarly, $a^{s_{l}}b^{s_{2}+1}c^{s_{3}}$ and
$a^{s_{l}}b^{s_{2}}c^{s_{3}+1}$ are the types of received directions
due to the second and third transmitter, respectively. Taking the
union of all these directions, one can compute $\mathcal{T}_2$.
However, we can easily see that the set of directions formed by
$a^{s_{l}}b^{s_{2}}c^{s_{3}}$, where  $0\leq s_{i}\leq n$ for all
$i\in\{1,2,3\}$ includes $\mathcal{T}_2$ and can be used as an upper
bound on the number of received directions. This set has $(n+1)^3$,
which is an upper bound for $L_2$. Hence, we conclude that
$$\eta=\frac{L_1}{L_2}>\left(\frac{n}{n+1}\right)^3.$$ Since $n$ is
an arbitrary integer, any alignment efficiency close to 1 is
possible. Hence, the  partial alignment approaches the perfect
alignment.

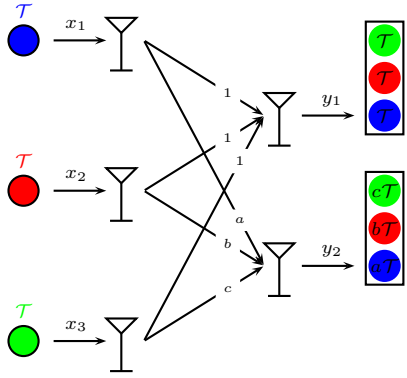
\begin{figure}
\centering
\begin{pspicture}(-1.8,-2.2)(3.8,2.8)
\rput(-.3,1.9){\antenna}
\rput(1.8,.9){\antenna}
\rput(1.8,-1.1){\antenna}
\rput(-.3,-0.1){\antenna}
\rput(-.3,-2.1){\antenna}

\rput(-1,2.3){\psline{->}(.7,0) \rput(.3,.2){\scriptsize $x_1$}}
\rput(-1,0.3){\psline{->}(.7,0)\rput(.3,.2){\scriptsize $x_2$}}
\rput(-1,-1.7){\psline{->}(.7,0)\rput(.3,.2){\scriptsize $x_3$}}
\rput{180}(3,1.3){\psline{<-}(.7,0)\rput{180}(.3,-.2){\scriptsize $y_1$}}
\rput{180}(3,-.7){\psline{<-}(.7,0)\rput{180}(.3,-.2){\scriptsize $y_2$}}

\cnode(0.2,2.3){0}{T1}
\cnode(0.2,0.3){0}{T2}
\cnode(0.2,-1.7){0}{T3}
\cnode(1.8,1.3){0}{R1}
\cnode(1.8,-0.7){0}{R2}

\ncline{->}{T1}{R1}
\ncput*[npos=.7]{\tiny{$1$}}
\ncline{->}{T2}{R1}
\ncput*[npos=.7]{\tiny{$1$}}
\ncline{->}{T3}{R1}
\ncput*[npos=.8]{\tiny{$1$}}
\ncline{->}{T1}{R2}
\ncput*[npos=.8]{\tiny{$a$}}
\ncline{->}{T2}{R2}
\ncput*[npos=.7]{\tiny{$b$}}
\ncline{->}{T3}{R2}
\ncput*[npos=.7]{\tiny{$c$}}

\def\mypoint #1#2#3{\pscircle[linecolor=#3,fillcolor=#3,fillstyle=solid](#1){.2}
\rput(#1){\scriptsize $\textcolor{black}{#2}$}}

\def\mymultipoint #1#2{\rput(#1){
\psframe(-.25,-.25)(.25,1.25)
\mypoint{0,0}{1}{#2}
\mypoint{0,.5}{2}{#2}
\mypoint{0,1}{3}{#2}
}}

\pscircle[fillcolor=red,fillstyle=solid](-1.4,.3){.2}
\rput(-1.4,.7){\scriptsize $\textcolor{red}{\mathcal{T}}$}
\pscircle[fillcolor=blue,fillstyle=solid](-1.4,2.3){.2}
\rput(-1.4,2.7){\scriptsize $\textcolor{blue}{\mathcal{T}}$}
\pscircle[fillcolor=green,fillstyle=solid](-1.4,-1.7){.2}
\rput(-1.4,-1.3){\scriptsize $\textcolor{green}{\mathcal{T}}$}

\rput(3.4,-.7){
\psframe(-.25,-.25)(.25,1.25)
\mypoint{0,0}{a\mathcal{T}}{blue}
\mypoint{0,.5}{b\mathcal{T}}{red}
\mypoint{0,1}{c\mathcal{T}}{green}
}

\rput(3.4,1.3){
\psframe(-.25,-.25)(.25,1.25)
\mypoint{0,0}{\mathcal{T}}{blue}
\mypoint{0,.5}{\mathcal{T}}{red}
\mypoint{0,1}{\mathcal{T}}{green}
}

\end{pspicture}
\caption{Three transmitters wish to align their signals at two
receivers. Each circle on the transmitters' sides represents a set
of data streams transmitted in the directions $\mathcal{T}$. Each
circle on the receivers' sides represent the received directions due
to different transmitters. The received directions can be aligned
with efficiency arbitrary close to one.}\label{fig alignment at two}
\end{figure}
\end{example}

For the multiple transmitter and receiver, the above approach can be
easily extended. In fact, it can be shown that the perfect alignment
is possible for any finite number of transmitters and receivers.

\mydef{Simultaneous Interference Alignment}{Partial interference
alignment of any finite number of signals is possible at any finite
number of receivers. Moreover, by increasing the number of transmit
directions, one can achieve any alignment efficiency close to one.}

The above example shows that a set of data streams with the
appropriate directions in any system can play the role of a single
data stream in the two-user $X$ channel, where perfect interference
alignment was possible. In addition to perfect alignment, which is
desired in any system, receivers are required to decode their own
messages from the received signals. However, the receiver can decode
its own messages if the intersection of the set of received
directions due to interference and message is null.

In the last example, we will combine all ideas presented in this
section to obtain the total DOF of the $3\times 3$ $X$ channel.

\begin{example}[$3\times 3$ $X$ channel]
In this example, we consider the $3\times 3$ $X$ channel shown in
Figure \ref{fig 3by3 x channel}. In this channel, each transmitter
has independent messages for all three receivers. Let $m_{ji}$
denote the message transmitted by the $i$th transmitter and intended
for the $j$th receiver, where $i,j\in\{1,2,3\}$. In addition, let
$x_{ij}$ denote the signal carrying the message $m_{ji}$.

The transmitters send their messages using the following signaling:
\begin{IEEEeqnarray*}{rl}
x_1=&h_{11}x_{11}+h_{21}x_{21}+h_{31}x_{31},\\
x_2=&h_{12}x_{12}+h_{22}x_{22}+h_{32}x_{32},\\
x_3=&h_{13}x_{13}+h_{23}x_{23}+h_{33}x_{33}.
\end{IEEEeqnarray*}

The messages intended for the first receiver are transmitted by
$x_{11}$, $x_{12}$, and $x_{13}$ (red circles in Figure \ref{fig
3by3 x channel}). In the previous example, we have shown that the
signals carrying these data can be efficiently aligned  at the
second and third receivers using transmit directions $\mathcal{G}_1$
generated by
$$G_1=\{h_{11}h_{21},h_{11}h_{31},h_{12}h_{22},h_{12}h_{32},h_{13}h_{23},h_{13}h_{33}\}.$$
To see this, let us consider the signal $x_{11}$. This signal
arrives at the second and third receivers multiplied by
$h_{11}h_{21}$ and $h_{11}h_{31}$, respectively. In other words,
$h_{11}h_{21}$ and $h_{11}h_{31}$ are equivalent channel gains
between $x_{11}$ and the second and third receivers. Therefore,
these factors need to be incorporated in the selection of transmit
directions to have efficient alignment at both receivers. A similar
argument can be applied for $x_{12}$ and $x_{13}$.   In a similar
fashion, one can obtain the sets $\mathcal{G}_2$ and $\mathcal{G}_3$
used for sending messages to the second and third receivers by using
generators
$$G_2=\{h_{21}h_{11},h_{21}h_{31},h_{22}h_{12},h_{22}h_{32},h_{23}h_{13},h_{23}h_{33}\}$$
and
$$G_3=\{h_{31}h_{11},h_{31}h_{21},h_{32}h_{12},h_{32}h_{22},h_{33}h_{13},h_{33}h_{23}\},$$
respectively.

The previous example ensures us that the preceding signaling is
efficient regarding interference alignment. However, we need to
guarantee that the messages can be decoded at the intended
receivers.  To this end, we look at the received directions at the
first receiver. $h_{11}^2\mathcal{G}_1$, $h_{12}^2 \mathcal{G}_1$,
and $h_{13}^2 \mathcal{G}_1$ are received directions due to the
intended messages. Clearly, they are all different and therefore can
be separated based on the separability condition. Moreover, it can
be shown that the set of intended directions has no intersection
with the set of interfering directions represented by $
\mathcal{G}_2\cup  \mathcal{G}_3$ (recall that $ \mathcal{G}_2$ and
$ \mathcal{G}_3$ are closed under multiplication). Dividing the
dimension of the first receiver into five, one can conclude that two
of them are occupied by interference and three of them are occupied
by the intended signals. Therefore, $\frac{3}{5}$ is an achievable
DOF at the first receiver. Due to symmetry, a similar argument can
be applied for the second and third receivers, resulting in
$\frac{9}{5}$ as the total DOF of the channel.

\begin{figure*}
\centering
\begin{pspicture}(-3.4,-2.2)(5.8,3.5)
\rput(-.9,1.9){\antenna}
\rput(-.9,-0.1){\antenna}
\rput(-.9,-2.1){\antenna}
\rput(1.8,1.9){\antenna}
\rput(1.8,-0.1){\antenna}
\rput(1.8,-2.1){\antenna}

\rput(-1.6,2.3){\psline{->}(.7,0) \rput(.3,.2){\scriptsize $x_1$}}
\rput(-1.6,0.3){\psline{->}(.7,0)\rput(.3,.2){\scriptsize $x_2$}}
\rput(-1.6,-1.7){\psline{->}(.7,0)\rput(.3,.2){\scriptsize $x_3$}}
\rput{180}(3,2.3){\psline{<-}(.7,0)\rput{180}(.3,-.2){\scriptsize $y_1$}}
\rput{180}(3,.3){\psline{<-}(.7,0)\rput{180}(.3,-.2){\scriptsize $y_2$}}
\rput{180}(3,-1.7){\psline{<-}(.7,0)\rput{180}(.3,-.2){\scriptsize $y_3$}}

\cnode(-0.4,2.3){0}{T1}
\cnode(-0.4,0.3){0}{T2}
\cnode(-0.4,-1.7){0}{T3}
\cnode(1.8,2.3){0}{R1}
\cnode(1.8,.3){0}{R2}
\cnode(1.8,-1.7){0}{R3}

\ncline{->}{T1}{R1}
\ncput*[npos=.7]{\tiny{$h_{11}$}}
\ncline{->}{T2}{R1}
\ncput*[nrot=:U,npos=.7]{\tiny{$h_{12}$}}
\ncline{->}{T3}{R1}
\ncput*[nrot=:U,npos=.75]{\tiny{$h_{13}$}}
\ncline{->}{T1}{R2}
\ncput*[nrot=:U,npos=.8]{\tiny{$h_{21}$}}
\ncline{->}{T2}{R2}
\ncput*[nrot=:U,npos=.7]{\tiny{$h_{22}$}}
\ncline{->}{T3}{R2}
\ncput*[nrot=:U,npos=.8]{\tiny{$h_{23}$}}
\ncline{->}{T1}{R3}
\ncput*[nrot=:U,npos=.8]{\tiny{$h_{31}$}}
\ncline{->}{T2}{R3}
\ncput*[nrot=:U,npos=.62]{\tiny{$h_{32}$}}
\ncline{->}{T3}{R3}
\ncput*[nrot=:U,npos=.7]{\tiny{$h_{33}$}}

\def\mypoint #1#2#3{\pscircle[linecolor=#3,fillcolor=#3,fillstyle=solid](#1){.2}
\rput(#1){\scriptsize $\textcolor{black}{#2}$}}

\def\mymultipoint #1#2{\rput(#1){
\psframe(-.25,-.25)(.25,1.25)
\mypoint{0,0}{1}{#2}
\mypoint{0,.5}{2}{#2}
\mypoint{0,1}{3}{#2}
}}

\mypoint{-2,2.3}{1}{red}
\mypoint{-2,.3}{2}{red}
\mypoint{-2,-1.7}{3}{red}

\mymultipoint{3.4,.3}{red}
\mymultipoint{3.4,-1.7}{red}

\mypoint{-2.6,2.3}{1}{blue}
\mypoint{-2.6,.3}{2}{blue}
\mypoint{-2.6,-1.7}{3}{blue}

\mymultipoint{3.4,2.3}{blue}
\mymultipoint{4,-1.7}{blue}

\mypoint{-3.2,2.3}{1}{green}
\mypoint{-3.2,.3}{2}{green}
\mypoint{-3.2,-1.7}{3}{green}

\mymultipoint{4,2.3}{green}
\mymultipoint{4,.3}{green}

\mypoint{4.6,2.3}{1}{red}
\mypoint{5.2,2.3}{2}{red}
\mypoint{5.8,2.3}{3}{red}

\mypoint{4.6,.3}{1}{blue}
\mypoint{5.2,.3}{2}{blue}
\mypoint{5.8,.3}{3}{blue}

\mypoint{4.6,-1.7}{1}{green}
\mypoint{5.2,-1.7}{2}{green}
\mypoint{5.8,-1.7}{3}{green}

\end{pspicture}
\caption{The $3\times 3$ $X$ channel. Each red circle represents a
set of data streams intended for the first receiver. They can be
aligned with efficiency arbitrary close to one at the second and
third receivers, occupying $\frac{1}{5}$ of the received directions.
Similarly, signals intended for the second and third receivers can
be aligned efficiently at non-intended receivers (blue and green
circles). Each set of data stream can carry data with multiplexing
gain of $\frac{1}{5}$. There are 9 sets of streams resulting in the
total DOF of $\frac{9}{5}$.}\label{fig 3by3 x channel}
\end{figure*}
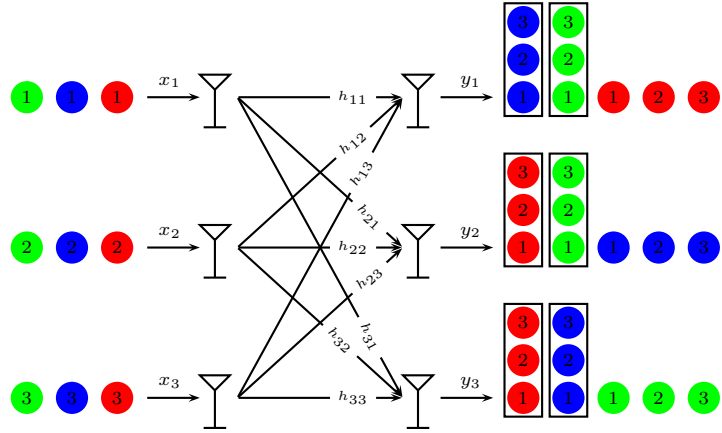
\end{example}


\section{Diophantine Approximation: Khintchine-Groshev Type Theorems}\label{sec groshev}
In number theory, the field of Diophantine approximation  deals with
the approximation of real numbers with rational numbers. The reader
is referred to \cite{schmidt,hardy} and the references therein.
Khintchine theorem is one of the cornerstones in this field. It
gives a criteria for a given function
$\psi:\mathbb{N}\to\mathbb{R}_+$ and real number $v$ such that $|p+v
q|<\psi(|q|)$  has either infinitely many solutions or at most
finitely many solutions for $(p,q)\in \mathbb{Z}^2$. Let
$\mathcal{A}(\psi)$ denote the set of real numbers such that $|p+v
q|<\psi(|v|)$  has infinitely many solutions in integers. The
theorem has two parts. The first part is the convergence part and
states that if $\psi(|q|)$ is convergent, i.e., $$ \sum_{q=1}^\infty
\psi(q)<\infty$$ then $\mathcal{A}(\psi)$ has measure zero with
respect to Lebesque measure. This part can be rephrased in a more
convenient way as follows. For almost all real numbers, $|p+v
q|>\psi(|q|)$ holds for all $(p,q)\in \mathbb{Z}^2$ except for
finitely many of them. Since the number of integers violating the
inequality is finite, one can find a constant $\kappa$ such that
$$|p+v q|>\kappa\psi(|q|)$$ holds for all integers $p$ and $q$
almost surely. The divergence part of the theorem states that
$\mathcal{A}(\psi)$ has the full measure, i.e. the set
$\mathbb{R}-\mathcal{A}(\psi)$ has measure zero, provided $\psi$ is
decreasing and $\psi(|q|)$ is divergent, i.e., $$ \sum_{q=1}^\infty
\psi(q)=\infty.$$

There is an extension to Khintchine's theorem due to Groshev, which
regards the rational approximation of linear forms with rational
coefficients. Let $\mathbf{v}=(v_1,v_2,\ldots,v_m)$ and
$\mathbf{q}=(q_1,q_2,\ldots,q_m)$ denote an $m$-tuple in
$\mathbb{R}^m$ and $\mathbb{Z}^m$, respectively. Let
$\mathcal{A}_m(\psi)$ denote the set of $m$-tuple real numbers
$\boldsymbol{g}$ such that
\begin{equation}
 |p+\mathbf{v}\cdot \mathbf{q}|<\psi (|\mathbf{q}|_{\infty})
\end{equation}
has infinitely many solutions for $p\in \mathbb{Z}$ and
$\mathbf{q}\in\mathbb{Z}^m$. $|\mathbf{q}|_{\infty}$ is the supreme
norm of $\mathbf{q}$ defined as $\max_i |q_i|$. The following
theorem gives the Lebesque measure of the set $\mathcal{A}_m(\psi)$.

\begin{theorem}[Khintchine-Groshev]\label{thm khintchine groshev}
Let $\psi:\mathbb{N}\to\mathbb{R}^+$. Then the set $\mathcal{A}_{m}(\psi)$ has measure zero, provided
\begin{equation}\label{convergence}
 \sum_{q=1}^\infty q^{m-1}\psi(q)<\infty,
\end{equation}
and has the full measure if
\begin{equation}
 \sum_{q=1}^\infty q^{m-1}\psi(q)=\infty \quad  \text{ and $\psi$ is monotonic}.
\end{equation}
\end{theorem}

In \cite{abolfazl-real}, Theorem \ref{thm khintchine groshev} is
used to prove that the total DOF of the two-user $X$ channel can be
achieved using a simple coding scheme. It is also proved that the
three-user GIC can achieve the DOF of $\frac{4}{3}$ almost surely.
Note that Theorem~\ref{thm khintchine groshev} does not include the
case where elements of $\mathbf{v}$ are related. It turned out that
such a shortcoming in this theorem prevented us from proving the
achievablity of $\frac{3}{2}$ for the three-user GIC.
Let us assume $\boldsymbol{v}$ lies on a manifold with dimension less than $m$ in
$\mathbb{R}^m$. In this case, the theorem may not be correct as the measure of the manifold is zero with respect to
Lebesque measure. Recently, \cite{groshev-berink} and \cite{groshev-Beresnevich} independently extended the
convergence part of the theorem to the class of non-degenerate manifolds. However, a subclass of non-degenerate
manifolds is sufficient for the proofs of the results in this paper.  Therefore, in the following theorem we state
the theorem in its simplest form by limiting the scope of it.

\begin{theorem}[\cite{groshev-berink} and \cite{groshev-Beresnevich}]\label{groshev-type}
Let $n\leq m$, $\mathbf{v}=(v_1,v_2,\ldots,v_n)\in \mathbb{R}^n$, and $g_1,g_2,\ldots, g_m$ be functions from
$\mathbb{R}^n$ to $\mathbb{R}$ with the following conditions:
\begin{enumerate}
\item $g_i$ for $i\in\{1,2,\ldots,m\}$ is analytic,
\item $1,g_1,g_2,\ldots, g_m$ are linearly independent over $\mathbb{R}$.
\end{enumerate}
For any monotonic function $\psi:\mathbb{N}\rightarrow\mathbb{R}_+$ such that $\sum_{q=1}^\infty
q^{m-1}\psi(q)<\infty$ the inequality
\begin{equation}
|p+q_1g_1(\mathbf{v})+q_2g_2(\mathbf{v})+\ldots+q_mg_m(\mathbf{v})|<\psi(|\mathbf{q}|_{\infty})
\end{equation}
has at most finitely many solutions $(p,\mathbf{q})\in \mathbb{Z}\times\mathbb{Z}^m$ for almost all
$\mathbf{v}\in\mathbb{R}^n$.
\end{theorem}

Throughout this paper, the function $\psi(q)$ is chosen as
$\frac{1}{q^{m+\epsilon}}$ for an arbitrary $\epsilon>0$. Clearly,
this function satisfies (\ref{convergence}) and is an appropriate
candidate for the theorem. If all conditions of the theorem hold,
then one can find a constant $\kappa$ such that for almost all
$\mathbf{v}\in\mathbb{R}^n$
\begin{equation}\label{khintchine}
 |p+q_1g_1(\mathbf{v})+q_2g_2(\mathbf{v})+\ldots+q_mg_m(\mathbf{v})|>\frac{\kappa}{(\max_i|q_i|)^{m+\epsilon}}
\end{equation}
holds for all $p\in \mathbb{Z}$ and $\mathbf{q}\in\mathbb{Z}^m$.

One class of functions satisfying the conditions in Theorem
\ref{groshev-type} is of special interest. Let
$\mathcal{G}(\mathbf{v})$ denote the set of all monomials with
variables from the set $\mathbf{v}=\{v_1,v_2,\ldots,v_n\}$. In other
words, a function $g$ belongs to $\mathcal{G}(\mathbf{v})$ if it can
be represented as $g=v_1^{s_1}v_2^{s_2}\cdots v_n^{s_m}$ for some
nonnegative integers $s_1,s_2,\ldots,s_n$. It is easy to show that
any collection of functions from $\mathcal{G}(\mathbf{v})$ satisfies
the conditions of Theorem \ref{groshev-type}. More specifically, all
functions belonging to $\mathcal{G}(\mathbf{v})$ are analytic.
Moreover, a set of monomials are independent over $\mathbb{R}$ as
long as they are distinct. As a special case when set $\mathbf{v}$
has only one member, i.e. $\mathbf{v}=\{v\}$, then we have
$\mathcal{G}(v)=\{1,v,v^2,v^3,\ldots\}$.

\section{Coding Scheme and Performance Analysis}\label{sec coding1}
Unless otherwise stated, we assume that the same encoding and
decoding  schemes are applied at all transmitters and all receivers,
respectively. In the following, we will describe the proposed
encoding and decoding schemes for a given transmitter and receiver.

\textbf{Construction of a single data stream}: Let us first explain
the encoding of a single data stream. The constellation
$\mathcal{C}=(-Q,Q)_{\mathbb{Z}}$ as the set of input symbols is
chosen. Even though one can use the continuum of real numbers as the
input alphabets, restriction to a finite set has the benefit of easy
and feasible interference alignment. We assume $Q=\gamma
P^{\frac{1-\epsilon}{2(m+\epsilon)}}$ where $\gamma$ is a constant.
Notice that since the number of input symbols are bounded by $2Q-1$,
the data stream modulated by $\mathcal{C}$ can at most provide
$\frac{1-\epsilon}{m+\epsilon}$ DOF. We will show that at high SNR
regimes this DOF can be achieved.

Having formed the constellation, a random codebook with rate $R$ is
constructed to change the channel into a reliable one. This can be
accomplished by choosing a probability distribution on the input
alphabets. The uniform distribution is the first candidate and it is
selected for the sake of simplicity. Note that since the
constellation is symmetrical by assumption, the expectation of the
uniform distribution is zero and the transmit signal has no DC
component. The power consumed by the data stream  can be loosely
upper-bounded as $Q^2$.

\begin{remark}
The parameters involved in the proposed construction, i.e, $Q$,
$m$,$\gamma$, and $\epsilon$, are universal and applied to all the
available data streams. Clearly, the optimum performance of a system
can be attained by selecting these parameters appropriately.
\end{remark}

\textbf{Encoding scheme}: It is well known that a transmitter with
average power constraint $P$ and equipped with $M$ antennas has $M$
degrees-of-freedoms available for data transmission. This is due to
the fact that the input signal lies on an $M$-dimensional Euclidean
space\footnote{If a channel is time/frequency varying then the input
signal over $M$ extensions of time/frequency lies on an
$M$-dimensional Euclidean space. Therefore, MIMO and time/frequency
varying channels behave the same regarding the DOF.}. As it has been
reported in numerous papers, the most applicable approach utilizing
these available DOFs relies on the expansion of the input signal
into $M$ bases and  transmission of a single data stream over each
of these bases. For instance, if the input signal is
$\mathbf{x}\in\mathbb{R}^M$ then by choosing
$\{\mathbf{T}_0,\mathbf{T}_2,\ldots,\mathbf{T}_{M-1}\}$ independent
vectors we have
\begin{equation}
\mathbf{x}=\sum_{i=0}^{M-1} \mathbf{T}_i x_i,
\end{equation}
where $x_i$ is the $i$'th component of $\mathbf{x}$ in the direction
of $\mathbf{T}_i$. Being a real number, $x_i$ for
$i\in\{0,1,\ldots,M-1\}$ can carry at most one DOF. If a
transmitter, however, wishes to send less than $M$ data streams, say
$L$, then it chooses $L$ bases out of $M$ available bases and
discards the rest of bases. In this scheme only integral DOFs are
possible for each transmitter. As a simple example, a single antenna
transmitter can send a data stream with either one DOF or zero DOF.

In this paper, we prove that the restriction on achieving integral
DOFs can be relaxed in a dramatic way. We claim that under some
regularity conditions, which are not too restrictive, any fractional
DOF is possible. Let us focus on a single antenna transmitter.
Viewing as a one-dimensional Euclidean space it has only one base;
whereas viewing as a vector field over rational numbers (or
equivalently integers), it has infinitely many bases.

The $i$'th transmitter chooses a set of real numbers, say
$\mathcal{T}_i=\{T_{i0},T_{i1},\ldots,T_{i(M-1)}\}$, as the set of
transmit directions for transmitting $M$ independent data streams.
The members of $\mathcal{T}_i$ are independent over the field of
rational numbers. In the proposed coding scheme, the transmit signal
can be represented by
\begin{equation}
x_i=A\sum_{l=0}^{M-1} T_{il} u_{il},
\end{equation}
where $u_{il}$ for $l\in\{0,1,\ldots,M-1\}$ is the $l$'th data
stream transmitted in the direction of $T_{il}$.

The parameter $A$ controls the input power of the transmitters. In
what follows, $A$ is computed based on an upper-bound on the input
power of a typical transmitter. We start with the following chain of
inequalities
\begin{IEEEeqnarray}{rl}
E[x_i^2]~&\stackrel{(a)}{=}A^2 \sum_{l=0}^{M-1}T_{il}^2 E\left[u_{il}^2\right]\nonumber\\
&\stackrel{(b)}{\leq} A^2 Q^2 \left(\sum_{l=0}^{M-1} T_{il}^2\right)\nonumber\\
&\stackrel{}{=} A^2 Q^2 \lambda_i^2 \nonumber
\end{IEEEeqnarray}
where (a) follows from the fact that all data streams are
independent and (b) follows from the fact that the data streams are
all the same and hence $u_{il}^2\leq Q^2$. We use a short-hand
notation $\lambda_i$ as $\lambda_i=\sum_{l=0}^{M-1} T_{il}^2$. Since
each $T_{il}$ is constant, $\lambda_i$ is also a constant. To
satisfy the power constraint, it is required that $$A\leq
\frac{P^{\frac{1}{2}}}{Q\lambda_i}$$ for all $i\in\{1,2,\ldots,K\}$,
where $K$ is the number of transmitters. Clearly, it is sufficient
to choose $$A= \frac{\zeta P^{\frac{1}{2}}}{Q}$$ where
$\zeta=\min_{i}\frac{1}{\lambda_i}$. By assumption $Q=\gamma
P^{\frac{1-\epsilon}{2(m+\epsilon)}}$. Hence, we have
\begin{equation}
A=\xi P^{\frac{m-1+2\epsilon}{2(m+\epsilon)}},
\end{equation}
where $\xi=\frac{\zeta}{\gamma}$.

In fact, $A$ and $Q$ are two important design parameters in the
encoding. $Q$ controls the cardinality of the input constellation,
which in turn provides the maximum achievable rate for individual
data streams. The cardinality of the constellation grows roughly as
$P^{\frac{1}{2m}}$.  On the other hand, $A$ controls the minimum
distance in the received constellation, which in turn affects the
performance. Our calculation reveals that no matter how many data
streams each transmitter is intended to send,  $Q$ and $A$ only
depend on $m$, which is the reciprocal of the multiplexing gain of
each data streams.

\textbf{Transmit directions and interference alignment}: The most
important part of the proposed coding design is the selection of
transmit directions. As it is shown in Section \ref{sec main ideas}
through several examples, transmit directions provide interference
alignment as well as separability at all receivers. In fact, to
design the optimum directions, interference alignment plays the role
as the separability condition usually comes for free.

One important observation is that the transmit directions need to be
generated based on channel parameters. Monomials are the best
candidates as they are forming a non-degenerate manifold in higher
dimensions. This property allows us to incorporate
Khintchine-Groshev type theorems in the performance analysis.

We will explain in more details how the transmit directions can be
chosen based on the channel coefficients. To have a glimpse on the
procedure, we consider there are $K'$ receivers in the network
receiving signals from $K$ transmitters as interference. The
interference, due to all transmitters at Receiver $j$, can be
represented as $I_j=\sum_{i=1}^{K} h_{ji}x_i$ where $h_{ji}$ is a
real number. Clearly, if each transmitter transmits one data stream
then it is impossible to align them at all receivers. However, it is
possible to align the transmit signals if each transmitter encodes
$M$ data streams as $x_i=\sum_{l=0}^{M-1} T_{il}u_{il}$ for all
$i\in\{1,2,\ldots,K\}$. Therefore, the interference at the $j$'th
receiver can be written as $$I_j=\sum_{i=1}^{K} \sum_{l=0}^{M-1}
(h_{ji}T_{il})u_{il},$$ where $(h_{ji}T_{il})$ is the \emph{received
direction} for the $l$'th data stream of the $i$'th transmitter. If
the transmit directions are chosen randomly, then all of the
received directions are distinct and the total number of received
directions is $KK'$, which is not desirable.

To reduce the number of received directions, $h_{ji}$'s can be used
as the generators for the transmit directions. Let us fix the set of
received directions by assuming that all received directions belong
to the set $\mathcal{T}_r$, which consists of directions of the form
$\prod_{i=1}^{K}\prod_{j=1}^{K'} h_{ji}^{s_{ji}}$, where $n$ is an
arbitrary integer and $0\leq s_{ji}\leq n$ for all
$i\in\{1,2,\ldots,K\}$ and $j\in\{1,2,\ldots,K'\}$. If $M'$ denotes
the number of received directions then it is easy to show that
$M'=(n+1)^{KK'}$.

A transmit direction is legitimate if it arrives at all receivers
with directions belonging to $\mathcal{T}_r$. Let us focus on the
$i$'th transmitter. The received signals due to the transmit signal
$x_i$ are $h_{1i}x_i,h_{2i}x_i,\ldots,h_{Mi}x_i$. If we choose the
transmit directions from the set $\mathcal{T}_i$, which consists of
directions of the form
$$\underbrace{\left(\prod_{j=1}^{K'} h_{ji}^{s'_{ji}}\right)}_{0\leq
s'_{ji}\leq n-1}\underbrace{\left(\prod_{k=1\& k\neq
i}^{K}\prod_{j=1}^{K'} h_{jk}^{s_{jk}}\right)}_{0\leq s_{jk}\leq
n},$$ then the received directions at all receivers belong to
$\mathcal{T}_r$ as the power of $h_{ji}$ for all
$j\in\{1,2,\ldots,K'\}$ is lowered by one. It is easy to show that
$M=n^{K'}(n+1)^{K'(K-1)}$.

The efficiency of the alignment can be measured by the ratio of $M'$
and $M$, i.e., $\eta=\frac{M'}{M}=\left(\frac{n+1}{n}\right)^{K'}$.
The perfect alignment happens when $M=M'$ , i.e., the ratio is one.
However, as $n$ can be chosen arbitrarily large, then we can have
any efficiency close to one from the proposed alignment technique.
In a loose sense, we can say that any number of transmitters can
align their signals at any number of receivers.

\textbf{Decoding scheme}: The received signal at the $j$'th receiver in its general form can be
represented by
\begin{equation}
y_j=A\left(\sum_{l=0}^{L_j-1}\bar{T}_{jl}u_{jl}+\sum_{l=0}^{L'_j-1}\bar{T}'_{jl}u'_{jl}\right)+z_j,
\end{equation}
where $\bar{T}_{jl}$ and $\bar{T}'_{jl}$ are the received directions
due to an intended data stream and an interfering signal,
respectively. It is assumed that $L_j$ and $L'_j$ are the number of
received directions due to the intended data streams and interfering
signals, respectively. $u_{jl}$ is an intended data stream.
$u'_{jl}$ is the an interfering signals. Because of interference
alignment, it is possible that $f_{jl}$ data streams arrive at the
direction $\bar{T}'_{jl}$, which results in $u'_{jl}\in
(-f_{jl}Q,f_{jl}Q)_{\mathbb{Z}}$. To have a uniform bound, let us
define $f=\max_{(j,l)} f_{jl}$ and
$\mathcal{U}'=(-fQ,fQ)_{\mathbb{Z}}$. Clearly, $u'_{jl}\in
\mathcal{U}'$ for all $j$'s and $l$'s.

We assume that $L_j+L'_j\leq m$ for all $j\in\{1,2,\ldots,K\}$. The
$j$'th receiver is interested in data streams $u_{jl}$ for all
$l\in\{0,1,\ldots,L_j-1\}$. The data stream $u_{jl}$ is decoded as
follows. The received signal is first passed through a hard decoder.
The hard decoder looks at the received constellation
$$\mathcal{V}_j=A\left(\sum_{l=0}^{L_j-1}\bar{T}_{jl}\mathcal{U}+\sum_{l=0}^{L'_j-1}\bar{T}'_{jl}\mathcal{U}'\right)$$ and maps
the received signal to the nearest point in the constellation. This changes the continuous channel to a discrete one
in which the input symbols are from the transmit constellation $\mathcal{U}$ and the output symbols are from the
received constellation $\mathcal{V}_j$.

It is assumed that the received constellation has the property that
there is a many-to-one map from $\mathcal{V}_{j}$ to
$\mathcal{U}_{j}=\sum_{l=0}^{L_j-1}\bar{T}_{jl}\mathcal{U}$. Recall
that the transmit directions are chosen in such a way that all
$u_{jl}$'s can be recovered uniquely from $\mathcal{U}_{j}$. This,
in fact, implies that if there is no additive noise in the channel
then the receiver can decode all intended data streams with zero
error probability. This property holds, for example, when
$\bar{T}_{jl}$'s and $\bar{T}'_{jl}$ are all distinct and linearly
independent over rational numbers. Throughout this paper, we always
design the transmit directions in such a way that this condition
holds.


The equivalent channel between $u_{jl}$ and the output of the hard
decoder $\hat{u}_{jl}$ becomes a discrete channel and  the
joint-typical decoder can be used to decode the data stream from a
block of $\hat{u}_{jl}$'s. To decode another data stream, Receiver
$j$ performs the same procedure used for decoding $u_{jl}$. In fact,
joint-decoding is not used to decode all intended data streams.

\textbf{Performance Analysis}: Let $d_{j_{\min}}$ denote the minimum
distance in the received constellation $\mathcal{V}_{j}$. The
average error probability in the equivalent discrete channel from
input $u_{jl}$ to output $\hat{u}_{jl}$ , i.e.
$P_e=Pr\{\hat{u}_{jl}\neq u_{jl}\}$ is bounded as:
\begin{IEEEeqnarray}{rl}\label{error probability}
P_e &\leq Q\left(\frac{d_{j_{\min}}}{2}\right)\leq
\exp\left({-\frac{ d_{j_{\min}}^2}{8}}\right).
\end{IEEEeqnarray}
$P_e$ can be used to lower bound the rate achievable for the data
stream $u_{jl}$. In \cite{Etkin-Ordentlich}, Etkin and Ordentlich
used Fano's inequality to obtain a lower bound on the achievable
rate, which is tight in high SNR regimes. Following similar steps,
one can obtain
\begin{IEEEeqnarray}{rl}
 R_{jl} & = I(\hat{u}_{jl},u_{jl})\nonumber\\
     & =H(u_{jl})-H(u_{jl}|\hat{u}_{jl})\nonumber\\
     & \stackrel{(a)}{\geq} H(u_{jl})-1-P_e\log |\mathcal{U}|\nonumber\\
     & \stackrel{(b)}{=} (1-P_e)\log |\mathcal{U}|-1 \nonumber\\
     & \stackrel{(c)}{=}(1-P_e)\log (2Q-1)-1\label{lower bound on R}
\end{IEEEeqnarray}
where (a) follows from Fano's inequality, (b) follows from the fact
that $u_{jl}$ has a uniform distribution on its range, and (c)
follows from the fact that $|\mathcal{U}|$, which is the number of
integers in the interval $[-Q,Q]$, is bounded by $2Q-1$. Let us
assume that $P_e\rightarrow 0$ as $P\rightarrow\infty$. Under this
condition, the achievable multiplexing gain from data stream
$u_{jl}$ can be obtained as follows:
\begin{IEEEeqnarray}{rl}
 r_{jl} &= \lim_{P\rightarrow \infty} \frac{R_{jl}}{0.5\log P}\nonumber\\
     & \geq \lim_{P\rightarrow \infty} \frac{\log Q}{0.5\log P}\nonumber\\
     & \stackrel{(a)}{=} \frac{1-\epsilon}{m+\epsilon}\label{Multiplexing gain}
\end{IEEEeqnarray}
where (a) follows from the fact that $Q=\gamma
P^{\frac{1-\epsilon}{2(m+\epsilon)}}$. Since $\epsilon>0$ is an
arbitrary constant, the multiplexing gain of $\frac{1}{m}$ is
achievable for the data stream $u_{jl}$.

Provided that all intended data streams can be successfully decoded
at all receivers, the achievable DOF at the $j$'th receiver can be
written as $\frac{L_j}{m}$. However, it is achievable under the
condition that $P_e\rightarrow 0$ as $P\rightarrow\infty$ and it
needs to be shown. To this end, one requires to calculate the
minimum distance between points in the received constellation.

Recall that $L_j+L'_j\leq m$ and $\bar{T}_{jl}$'s and
$\bar{T}'_{jl}$'s are all distinct and monomials with variables from
the channel coefficients. Theorem \ref{groshev-type} can be applied
to obtain a lower bound on the minimum distance. Let us  assume that
one of the directions in $\bar{T}_{jl}$'s or $\bar{T}'_{jl}$' is 1.
Then a point in $\mathcal{V}_j$ can be represented as
\begin{equation}
v=A\left(v_0+\sum_{l=1}^{L_j+L_j-1}\hat{T}_{l}v_{l}\right).
\end{equation}
where $\hat{T}_{l}$'s are all distinct monomials at receiver $j$.
Moreover, $v_l$ for all $l\in\{0,1,\ldots,L_j+L'_j-1\}$ are bounded
by $(-fQ,fQ)_{\mathbb{Z}}$. Therefore, the distance  between any two
points in the received constellation $\mathcal{V}_j$ can be bounded
using (\ref{khintchine}) as follows:
\begin{equation}\nonumber
d_{j_{\min}}>\frac{\kappa A}{(2fQ)^{L_j+L'_j-1+\epsilon}}.
\end{equation}
Since $L_j+L'_j\leq m$, we have
\begin{equation}\label{khintchine-mg}
d_{j_{\min}}>\frac{\kappa A}{(2fQ)^{m-1+\epsilon}}.
\end{equation}

The probability of error in hard decoding (see (\ref{error
probability})) can be bounded as
\begin{equation}\label{alaki1}
P_e<\exp\left({-\varrho\left(\frac{ A}{Q^{m-1+\epsilon}}\right)^2}\right),
\end{equation}
where $\eta$ is a constant and a function of $\gamma$, $\kappa$, $\sigma$, and $\gamma_i$s.

Substituting $A$ and $Q$ in (\ref{alaki1}) yields
\begin{equation}
P_e<\exp\left(- \eta P^{\epsilon}\right),
\end{equation}
which shows that $P_e$ has the desired property.

The following theorem summarizes the conditions needed to achieve
the multiplexing gain of $\frac{1}{m}$ per data stream.
\begin{theorem}\label{basic1}
Consider there are $K$ transmitters and $K'$ receivers in a system
parameterized by the channel coefficient vector $\mathbf{h}$.
Transmitter $i$ sends $M$ data stream along directions
$\mathcal{T}_i=\{T_{i0},T_{i2},\ldots,T_{i(M-1)}\}$ for all $i\in
\{1,2,\ldots,K\}$. The data streams intended for the $j$'th receiver
arrive at $L_j$ directions, which are
$\mathcal{T}_j=\{\bar{T}_{j0},\bar{T}_{j2},\ldots,\bar{T}_{j(L_j-1)}\}$.
Moreover, the interference part of the received signal at the $j$'th
receiver has $L'_j$ effective data streams with received directions
$\mathcal{T}'_j=\{\bar{T}'_{j0},\bar{T}'_{j2},\ldots,\bar{T}'_{j(L'_j-1)}\}$
for all $j\in \{1,2,\ldots,K'\}$. Let the following conditions for
all $j\in \{1,2,\ldots,K'\}$ hold:
\begin{description}
  \item[C1] Components of $\mathcal{T}_i$ are distinct member of $\mathcal{G}(\mathbf{h})$ and linearly independent over the field of rational numbers.
  \item[C2] Components of $\mathcal{T}_{i}$ and $\mathcal{T}'_{i}$ are all distinct.
  \item[C3] One of the elements of either $\mathcal{T}_{i}$ or $\mathcal{T}'_{i}$ is 1.
\end{description}
Then, by encoding each data stream using the constellation $\mathcal{U}=(-Q,Q)_{\mathbb{Z}}$ where $Q=\gamma
P^{\frac{1-\epsilon}{2(m+\epsilon)}}$ and $\gamma$ is a constant, the following DOF is achievable for almost all realizations of the system:
\begin{equation}
r_{\text{sum}}=\frac{L_1+L_2+\cdots+L_{K'}}{m},
\end{equation}
where $m$ is the maximum received directions among all receivers, i.e., $m=\max_i L_i+L'_i$.
\end{theorem}

\begin{remark}
If C2 holds, then the measure of the event ``components of
$\mathcal{T}_{i}$ and $\mathcal{T}'_{i}$ are dependent over the
field of rational numbers'' is zero.
\end{remark}
\begin{remark}
If C3 does not hold, then by adding a virtual data stream in the
direction 1 at the receiver, one can conclude that $\frac{1}{m+1}$
is achievable for all data streams.
\end{remark}

Theorem \ref{basic1} implies that the most difficult part of the
design is the selection of transmit directions for all users. This
is due to the fact that random selection results in $m=\sum_{i=1}^K
L_i$ received directions, which in turn provides 1 DOF for the
channel. A careful design is needed to reduce the number of received
directions at all users. In the following section, we provide such a
design for the $K$-user GIC.

\section{$K$-user Gaussian Interference Channel}

\subsection{System Model}\label{sec system model}

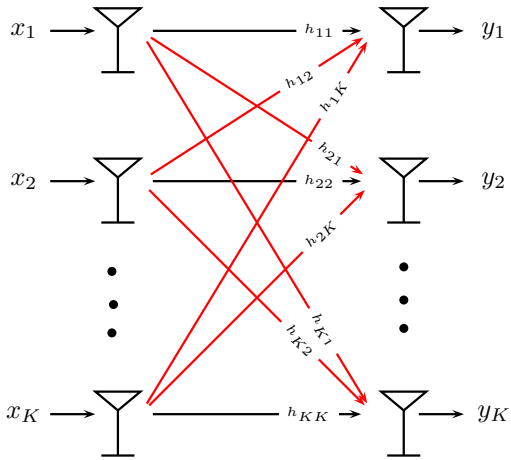
\begin{figure}
 \centering
\scalebox{1} 
{
\begin{pspicture}(0,-2.99)(6.4628124,2.97)
\psset{linewidth=0.03cm,arrowsize=0.05291667cm
2.0,arrowlength=1.4,arrowinset=0.4}
\usefont{T1}{ptm}{m}{n}

\def\antenna{%
\begin{pspicture}(1,1)
\pstriangle[gangle=-180.0](0,0.25)(0.6,0.25)
\psline(0,0)(0,-0.60)
\psline(-0.22,-0.60)(0.22,-0.60)
\end{pspicture}
}

\rput[bl](1.7,2.5){\antenna}
\rput[bl](1.7,0.5){\antenna}
\rput[bl](1.7,-2.6){\antenna}

\rput[bl](5.5,2.5){\antenna}
\rput[bl](5.5,0.5){\antenna}
\rput[bl](5.5,-2.6){\antenna}

\psline{->}(.8,2.45)(1.4,2.45)
\psline{->}(5.7,2.45)(6.3,2.45)
\rput(0.45,2.45){$x_1$}
\rput(6.7,2.45){$y_1$}

\psline{->}(.8,0.45)(1.4,0.45)
\psline{->}(5.7,0.45)(6.3,0.45)
\rput(0.45,0.45){$x_2$}
\rput(6.7,0.45){$y_2$}

\psline{->}(.8,-2.65)(1.4,-2.65)
\psline{->}(5.7,-2.65)(6.3,-2.65)
\rput(0.45,-2.65){$x_K$}
\rput(6.7,-2.65){$y_K$}

\cnode[linecolor=white](2,2.45){.15}{T1}
\cnode[linecolor=white](2,0.45){.15}{T2}
\cnode[linecolor=white](2,-2.65){.15}{Tk}

\cnode[linecolor=white](5.1,2.45){.15}{R1}
\cnode[linecolor=white](5.1,0.45){.15}{R2}
\cnode[linecolor=white](5.1,-2.65){.15}{Rk}

\ncline{->}{T1}{R1}
\ncput*[npos=.8]{\tiny $h_{11}$}
\ncline[linecolor=red]{->}{T1}{R2}
\ncput*[nrot=:U,npos=.85]{\tiny $h_{21}$}
\ncline[linecolor=red]{->}{T2}{R1}
\ncput*[nrot=:U,npos=.7]{\tiny $h_{12}$}
\ncline{->}{T2}{R2}
\ncput*[npos=.8]{\tiny $h_{22}$}
\ncline{->}{Tk}{Rk}
\ncput*[npos=.75]{\tiny $h_{KK}$}
\ncline[linecolor=red]{->}{Tk}{R1}
\ncput*[nrot=:U,npos=.85]{\tiny $h_{1K}$}
\ncline[linecolor=red]{->}{Tk}{R2}
\ncput*[nrot=:U,npos=.8]{\tiny $h_{2K}$}
\ncline[linecolor=red]{->}{T1}{Rk}
\ncput*[nrot=:U,npos=.8]{\tiny $h_{K1}$}
\ncline[linecolor=red]{->}{T2}{Rk}
\ncput*[nrot=:U,npos=.7]{\tiny $h_{K2}$}

\psdots[dotsize=0.12](5.5,-0.69)
\psdots[dotsize=0.12](5.5,-1.13)
\psdots[dotsize=0.12](5.5,-1.51)
\psdots[dotsize=0.12](1.6209375,-0.75)
\psdots[dotsize=0.12](1.6409374,-1.19)
\psdots[dotsize=0.12](1.6209375,-1.57)
\end{pspicture}
}
\caption{The $K$-user GIC. User $i$ for $i\in \{1,2,\ldots,K\}$ wishes to communicate with its
corresponding receiver while receiving interference from other users.}\label{k-user IC}
\end{figure}

The $K$-user GIC models a network in which $K$ transmitter-receiver
pairs (users) sharing a common bandwidth wish to have reliable
communication at their maximum rates. The channel's input-output
relation can be stated as follows, see Figure \ref{k-user IC},
\begin{IEEEeqnarray}{rl}\label{k-user model}
 y_1 &=h_{11}x_1+h_{12}x_2+\ldots +h_{1K}x_K+z_1,\nonumber\\
 y_2 &=h_{21}x_1+h_{22}x_2+\ldots +h_{2K}x_K+z_2,\nonumber\\
\vdots\ &=\quad  \vdots \quad\qquad \vdots \quad\qquad\ddots\qquad\vdots \\
y_K &=h_{K1}x_1+h_{K2}x_2+\ldots +h_{KK}x_K+z_K,\nonumber
\end{IEEEeqnarray}
where $x_i$ and $y_i$ are input and output symbols of User $i$ for
$i\in\{1,2,\ldots,K\}$, respectively. $z_i$ is Additive White Gaussian Noise (AWGN) with unit
variance for $i\in\{1,2,\ldots,K\}$. Transmitters are subject to the
power constraint $P$. $h_{ji}$ represents the channel gain between
Transmitter $i$ and Receiver $j$. It is assumed that all channel
gains are real and time invariant. The set of all channel gains is
denoted by $\mathbf{h}$, i.e.,
$\mathbf{h}=\{h_{11},\ldots,h_{1K},h_{21},\ldots,h_{2K},\ldots,h_{K1},\ldots,h_{KK}\}$.
Since the noise variances are normalized, the Signal to Noise Ratio
(SNR) is equivalent to the input power $P$. Hence, we use them
interchangeably throughout this paper.

In this paper, we are primarily interested in characterizing the
total DOF of the $K$-user GIC. Let $\mathcal{C}$ denote the capacity
region of this channel. The DOF region  associated with the channel
is in fact the shape of $\mathcal{C}$ in high SNR regimes scaled by
$\log \text{SNR}$. Let us denote the DOF region by $\mathcal{R}$.
All extreme points of $\mathcal{R}$ can be identified by solving the
following optimization problem:
\begin{equation}
r_{\boldsymbol{\lambda}}=\lim_{\text{SNR}\rightarrow\infty}\max_{\mathbf{R}\in
\mathcal{C}}\frac{\boldsymbol{\lambda}^t \mathbf{R}}{0.5\log
\text{SNR}}.
\end{equation}
The total DOF refers to the case where
$\boldsymbol{\lambda}=\{1,1,\ldots,1\}$, i.e., the sum-rate is
concerned. Throughout this paper, $r_{\text{sum}}$ denotes the total
DOF of the system.

An upper bound on the DOF of this channel is obtained in
\cite{cadambe2008iaa}. The upper bound states that the total DOF of
the channel is less than $\frac{K}{2}$, which means each user can at
most enjoy one half of its maximum DOF.

\subsection{Three-user Gaussian Interference Channel:  $\text{DOF}=\frac{3}{2}$ is Achievable}\label{sec three-user}
In this section, we consider the three-user GIC and  explain in
detail that, by an appropriate selection of transmit directions, the
DOF of $\frac{3}{2}$ is achievable for almost all cases. We will
explain in more detail that by an appropriate selection of transmit
directions this DOF can be achieved.

In \cite{abolfazl-real}, we defined the standard model of the three-user GIC. The
definition is as follows:
\begin{definition}
The three user interference channel is called standard if it can be represented as
\begin{IEEEeqnarray}{rl}\label{alaki7}
y_1&=G_1x_1+x_2+x_3+z_1\nonumber\\
y_2&=G_2x_2+x_1+x_3+z_2\\
y_3&=G_3x_3+x_1+G_0x_2+z_3.\nonumber
\end{IEEEeqnarray}
where $x_i$ for User $i$ is subject to the power constraint $P$.
$z_i$ at Receiver $i$ is AWGN with unit variance.
\end{definition}

In \cite{abolfazl-real}, it is also proved that every three-user GIC
has an equivalent standard channel as far as the DOF is concerned.
The parameters in the standard channel are related to the parameters
of the original one thorough the following equations.
\begin{IEEEeqnarray*}{rl}
&G_0=\frac{h_{13}h_{21}h_{32}}{h_{12}h_{23}h_{31}},\\
&G_1=\frac{h_{11}h_{12}h_{23}}{h_{12}h_{21}h_{13}},\\
&G_2=\frac{h_{22}h_{13}}{h_{12}h_{23}},\\
&G_3=\frac{h_{33}h_{12}h_{21}}{h_{12}h_{23}h_{31}}.
\end{IEEEeqnarray*}

As mentioned in the previous section, transmit directions are
monomials with variables from channel coefficients. For the three
user case, we only use $G_O$ as the generator of transmit
directions. Therefore, transmit directions are selected from the set
$\mathcal{G}(G_0)$, which is a subset of
$\mathcal{G}(G_0,G_1,G_2,G_3)$. Clearly,
$\mathcal{G}(G_0)=\{1,G_0,G_0^2,G_0^3,\cdots\}$.

We consider two different cases based on the value of $G_0$ being
algebraic or transcendental. Although the measure of being
algebraic is zero, we prove that for each case the total DOF can be
achieved if the transmit and receive directions satisfy the
conditions of Theorem \ref{basic1}. We start with the case where
$G_0$ is algebraic.

\subsubsection{Case I: $G_0$ is algebraic}
By definition, if $G_0$ is algebraic then it is a root of a
polynomial with integer coefficients. Let us assume $G_0$ satisfies
\begin{equation}\label{algebraic}
a_dG_0^d+a_{d-1}G_0^{d-1}+\ldots+a_1G_0+a_0=0,
\end{equation}
where $a_d,a_{d-1},\ldots,a_0$ are integers. In other words, the set
$\mathcal{T}=\{1,G_0,G_0^2,\ldots,G_0^{d-1}\}$ is a basis for
$\mathcal{G}(G_0)$ over rational numbers. Therefore, as the transmit
directions need to be independent over the field of rational
numbers, the transmitters are restricted to choose their transmit
directions among numbers in $\mathcal{T}$. We assume that all
transmitters transmit along all directions in $\mathcal{T}$, i.e.,
$\mathcal{T}_i=\mathcal{T}$ for all $i\in\{1,2,3\}$. By this
selection, C1 in Theorem \ref{basic1} holds for all transmitters.

In this case, Transmitter $i$ sends $L_i=d$ data streams as follows
\begin{equation}
x_i=A\sum_{j=0}^{d-1} G_0^j u_{ij},
\end{equation}
for all $i\in\{1,2,3\}$. The received signal at Receiver 1 can be written as
\begin{equation}
 y_1=A\left(\sum_{j=0}^{d-1} G_1G_0^j u_{1j}+\sum_{j=0}^{d-1} G_0^j u'_{1j}\right)+z_1,
\end{equation}
where $u'_{1j}=u_{2j}+u_{3j}$ for all $j\in\{0,1,\ldots,d-1\}$. The
signals from Transmitters 2 and 3 are aligned and the number of
received directions is $L'_1=d$. Moreover C2 and C3 in Theorem
\ref{basic1} hold for this receiver. Since the received signal at
Receiver 2 is similar to that of Receiver 1, we can deduce that
$L'_2=d$ and C2 and C3 hold.

The received signal at Receiver 3 can be written as
\begin{equation}\label{algebraic received}
y_3=A\left(\sum_{j=0}^{d-1} G_3G_0^j u_{3j}+\sum_{j=0}^{d} G_0^j u'_{3j}\right)+z_3,
\end{equation}
where $u'_{3j}=u_{2j}+u_{1(j-1)}$ for $j\in\{1,2,\ldots,d-1\}$,
$u'_{30}=u_{20}$, and $u'_{3d}=u_{1d}$. The number of received
directions from interfering users is $d+1$. However, they are not
independent over the field of rational numbers. Using
(\ref{algebraic}), $G_0^d$ can be represented as a linear
combination of $\{1,G_0,G_0^2,\ldots,G_0^{d-1}\}$ with rational
coefficients. Multiplying both sides of (\ref{algebraic received})
by $a_d$, we have
\begin{equation}
\tilde{y}_3=A\left(\sum_{j=0}^{d-1} a_dG_3G_0^j u_{3j}+\sum_{j=0}^{d-1} G_0^j a_d u'_{3j}+ a_d G_0^d u'_{3d}\right)+\tilde{z}_3,
\end{equation}
where $\tilde{y}_3=a_dy_3$ and $\tilde{z}_3=a_dz_3$. Substituting from (\ref{algebraic received}), we obtain
\begin{equation}
\tilde{y}_3=A\left(\sum_{j=0}^{d-1} a_dG_3G_0^j u_{3j}+\sum_{j=0}^{d-1} G_0^j (\underbrace{a_d u'_{3j}-a_j u'_{3d}}_{u''_j})\right)+\tilde{z}_3.
\end{equation}
Clearly, $L'_3=d$ and C2 and C3 hold for this receiver as well.

The maximum number of received directions at all receivers is
$m=2d$. Since C1, C2, and C3 hold at all receivers, by applying
Theorem \ref{basic1} we conclude that the total DOF of $\frac{3}{2}$
is achievable for almost all cases.

\begin{remark}
In a special case, $d=1$ in (\ref{algebraic}). In other words, $G_0$
is a rational number. This case is considered in
\cite{Etkin-Ordentlich} and it is proved that it can achieve the
total DOF of the channel.
\end{remark}

\subsubsection{Case II: $G_0$ is transcendental}
If $G_0$ is transcendental then all members of $\mathcal{G}(G_0)$
are linearly independent over the field of rational numbers. Hence,
we are not limited to any subset of $\mathcal{G}(G_0)$, as far as
the independence of transmit directions is concerned. We will show
that $\frac{3n+1}{2n+1}$ is an achievable DOF for any
$n\in\mathbb{N}$. To this end, we propose a design that is not
symmetrical.

Transmitter 1 uses the set of directions
$\mathcal{T}_1=\{1,G_0,G_0^2,\ldots,G_0^{n}\}$ to transmit $L_1=n+1$
to its corresponding receiver. Clearly $\mathcal{T}_1$ satisfies C1.
The transmit signal from User 1 can be written as
$$x_1=A\sum_{j=0}^n G_0^j u_{1j}.$$ Transmitters 2 and 3 transmit in
$L_2=L_3=n$ directions using
$\mathcal{T}_2=\mathcal{T}_3=\{1,G_0,G_0^2,\ldots,G_0^{n-1}\}$.
Clearly both $\mathcal{T}_2$ and $\mathcal{T}_3$ satisfy C1. The
transmit signals can be expressed as $$x_2=A\sum_{j=0}^{n-1} G_0^j
u_{2j}$$ and $$x_3=A\sum_{j=0}^{n-1} G_0^j u_{3j}.$$

The received signal at Receiver 1 can be expressed as:
\begin{equation}
 y_1=A\left(\sum_{j=0}^n G_1G_0^j u_{1j}+\sum_{j=0}^{n-1} G_0^j u'_{1j}\right)+z_1,
\end{equation}
where $u'_{1j}=u_{2j}+u_{3j}$. In fact, transmit signals from Users 2 and 3 are aligned at Receiver 1. This is
due to the fact that out of $2n$ possible received directions, only $n$ directions are effective, i.e.,
$L'_1=n$. One can also confirm that C2 and C3 hold at Receiver 1.

The received signal at Receiver 2 can be expressed as:
\begin{equation}
 y_2=A\left(\sum_{j=0}^{n-1} G_2G_0^j u_{2j}+\sum_{j=0}^{n} G_0^j u'_{2j}\right)+z_2,
\end{equation}
where $u'_{2j}=u_{1j}+u_{3j}$ for all $j\in\{0,1,\ldots,n-1\}$ and
$u'_{2n}=u_{1n}$. At Receiver 2, transmitted signals from Users 1
and 3 are aligned and the number of effective received directions is
$L'_2=n+1$. Moreover, it can be easily seen that C2 and C3 hold at
Receiver 2.

The received signal at Receiver 3 can be expressed as:
\begin{equation}
 y_3=A\left(\sum_{j=0}^{n-1} G_3G_0^j u_{3j}+\sum_{j=0}^{n} G_0^j u'_{3j}\right)+z_3,
\end{equation}
where $u'_{3j}=u_{1j}+u_{2j}$ for all $j\in\{1,2,\ldots,n\}$ and
$u'_{30}=u_{10}$. At Receiver 3, transmitted signals from Users 1
and 2 are aligned and the number of effective received directions is
$L'_2=n+1$. Clearly, C2 and C3 hold for Receiver 3.

Since C1, C2, and C3 hold at all users, we only need to obtain the
number of maximum received directions at all receivers. To this end,
we observe that $$m=\max\{L_1+L'_1,L_2+L'_2,L_3+L'_3\}=2n+1$$.
Therefore, an application of Theorem \ref{basic1} reveals that the
following DOF is achievable.
\begin{IEEEeqnarray}{rl}
r_{\text{sum}} & =\frac{L_1+L_2+L_3}{m}\nonumber\\
& =\frac{3n+1}{2n+1}.
\end{IEEEeqnarray}
Since $n$ is an arbitrary integer, one can conclude that
$\frac{3}{2}$ is achievable for the three-user GIC almost surely.

\subsection{$K$-user Gaussian Interference Channel: $\text{DOF}=\frac{K}{2}$ is Achievable}\label{sec k-user}
In this section, we prove the main theorem of this paper, i.e., the
DOF of $\frac{K}{2}$ is achievable for the $K$-user GIC. As pointed
out in Section \ref{sec coding1}, we need to design the transit
directions of all transmitters in such a way that they satisfy the
conditions of Theorem \ref{basic1}. Recall that all transmit
directions are monomials with variables in $\mathbf{h}$. We reserve
the direct gains and do not use them as generating variables. The
reason is that C2 in Theorem \ref{basic1} requires that all received
directions be distinct.  By setting aside the direct gains, a
transmit direction from the intended user is multiplied by the
direct gain and therefore it is distinct from all other transmit
directions (by C1 all transmit directions from a user are distinct).

We assume that all channel gains are transcendental. On the one
hand, since the measure of being algebraic is zero, this assumption
is innocuous. On the other hand, as we learned from the three-user
case algebraic gains are beneficial as they reduce the number of
transmit directions required to achieve the total DOF of the
channel.

We start with selecting the transmit directions for User $i$. A
direction $T\in\mathcal{G}(\mathbf{h})$ is chosen as the transmit
direction for User $i$ if it can be represented as
\begin{equation}\label{k-user transmit directions}
T=\prod_{j=1}^{K}\prod_{l=1}^{K}  h_{jl}^{s_{jl}},
\end{equation}
where $s_{jl}$'s are integers satisfying
\begin{equation}\nonumber
\begin{cases}
s_{jj}=0 & \forall ~ j\in\{1,2,\ldots,K\}\\
0\leq s_{ji}\leq n-1 &\forall ~ j\in\{1,2,\ldots,K\} ~\& ~j\neq i\\
0\leq s_{jl}\leq n & ~ \text{Otherwise.}
\end{cases}
\end{equation}
The set of all transmit directions is denoted by $\mathcal{T}_i$. It is easy to show that the cardinality of this set is \begin{equation}
L_i=n^{K-1}(n+1)^{(K-1)^2}.
\end{equation}
Clearly, $\mathcal{T}_i$ satisfies C1 for all
$i\in\{1,2,\ldots,K\}$.

To compute $L'_i$ (the number of independent received directions due
to interference), we investigate the effect of Transmitter $k$ on
Receiver $i$. Let us first define $\mathcal{T}_r$ as the set of
directions represented by (\ref{k-user transmit directions}) and
satisfying
\begin{equation}\label{k-user condition1}
\begin{cases}
s_{jj}=0 & \forall ~ j\in\{1,2,\ldots,K\}\\
0\leq s_{jl}\leq n & ~ \text{Otherwise.}
\end{cases}
\end{equation}
We claim that $\mathcal{T}_{ik}$, the set of received directions at
Receiver $i$ due to Transmitter $k$, is a subset of $\mathcal{T}_r$.
In fact, all transmit directions of Transmitter $k$ arrive at
Receiver $i$ multiplied by $h_{ik}$. Based on the selection of
transmit directions, however, the maximum power of $h_{ik}$ in all
members of $\mathcal{T}_{ik}$ is $n-1$. Therefore, none of the
received directions violates the condition of (\ref{k-user
condition1}) and this proves the claim.

Since $\mathcal{T}_r$ is not related to User $k$, one can conclude
that $\mathcal{T}_{ik}\subseteq \mathcal{T}_r$ for all
$k\in\{1,2,\ldots,K\}$ and $k\neq i$. Hence, we deduce that all
interfering users are aligned in the directions of $\mathcal{T}_r$.
Now, $L'_i$ can be obtained by counting the members of
$\mathcal{T}_r$. It is easy to show that
\begin{equation}
L'_i=(n+1)^{K(K-1)}.
\end{equation}

The received directions at Receiver $i$ are members of
$h_{ii}\mathcal{T}_i$ and $\mathcal{T}_r$. Since $h_{ii}$ does not
appear in members of $\mathcal{T}_r$, the members of
$h_{ii}\mathcal{T}_i$ and $\mathcal{T}_r$ are distinct. Therefore,
C2 holds at Receiver $i$. Since all the received directions are
irrationals, C3 does not  hold at Receiver $i$.

Since $C_1$ and $C_2$ hold for all users, we can apply Theorem \ref{basic1} to obtain the DOF of the channel. We have
\begin{IEEEeqnarray}{rl}
r_{\text{sum}}&=\frac{L_1+L_2+\ldots+L_K}{m+1}\nonumber\\
&=\frac{Kn^{K-1}(n+1)^{(K-1)^2}}{m+1}
\end{IEEEeqnarray}
where $m$ is
\begin{IEEEeqnarray}{rl}
 m &=\max_i L_i+L'_i\nonumber\\
   &=n^{K-1}(n+1)^{(K-1)^2}+(n+1)^{K(K-1)}.
\end{IEEEeqnarray}
Combining the two equations, we obtain
\begin{equation}
r_{\text{sum}}=\frac{K}{1+(\frac{n+1}{n})^{K-1}+\frac{1}{n^{K-1}(n+1)^{(K-1)^2}}}.
\end{equation}
Since $n$ can be arbitrary large, we conclude that $\frac{K}{2}$ is achievable for the $K$-user GIC.

\section{Cellular Systems: Uplink}\label{sec uplink}
\subsection{System Model}

In a cellular network, an area is partitioned into several cells
and, within each cell, there is a base station serving users inside
the cell. There are two modes of operation. In the uplink mode,
users within a cell transmit independent messages to the base
station in the cell; whereas in the downlink mode, the base station
broadcasts independent messages to all users inside the cell. In
this section, we only consider the uplink mode. Abstractly, the
uplink mode corresponds to a network in which several Multiple
Access Channels (MAC) share the same spectrum for data transmission.
Let us assume there exist $M$ users in each MAC and there are $K$
MACs in the network. The received signal at the base station in Cell
$k$ can be represented as
\begin{equation}
y_k=\underbrace{\sum_{l=1}^{M} h_{k(kl)}x_{kl}}_{\text{users within
the cell}}+\underbrace{\sum_{i=1 \& i\neq k}^{K}
I_{ki}}_{\text{intra cell interference}}+z_k
\end{equation}
where $I_{ki}$ is the aggregate interference from all users in Cell $i$, i.e.,
\begin{equation}
I_{ki}=\sum_{l=1}^{M} h_{k(il)} x_{il}.
\end{equation}

Let $\mathcal{C}_{\text{up}}$  denote the capacity region of this
channel. The DOF region associated with the channel can be defined
as the shape of the region in high SNR regimes scaled by $\log
\text{SNR}$. Let us denote the DOF region by
$\mathcal{R}_{\text{up}}$. We are primarily interested in the main
facet of the DOF region defined as:
\begin{equation}
r_{\text{up}}=\lim_{\text{SNR}\rightarrow\infty}\max_{\mathbf{R}\in
\mathcal{C}_{\text{up}}}\frac{\sum_{k=1}^{K}\sum_{l=1}^{M} R_{kl}}{0.5\log\text{SNR}},
\end{equation}
where $R_{kl}$ is an achievable rate for the $l$'th user in Cell $k$.

\subsection{The Total DOF of $\frac{KM}{M+1}$ is Achievable}
To obtain an upper bound on the total  DOF of this channel, we
assume that all users within a cell can cooperate. This cooperation
converts the uplink mode to a MISO $K$-user GIC with $M$ antennas at
the transmitters and one antenna at the receivers. An upper bound on
the DOF of the MISO $K$-user GIC  is obtained in \cite{Huang-jafar}.
The upper bound states that the total DOF of the channel is less
than $\frac{KM}{M+1}$. We will show that this DOF is achievable.

We start with selecting the transmit directions of the $m$'th user
in Cell $k$. A direction $T\in\mathcal{G}(\mathbf{H})$ ($\mathbf{H}$
is the set of all channel gains) is chosen as the transmit direction
for this user if it can be represented as
\begin{equation}\label{k-user transmit directions1}
T=\prod_{j=1}^{K}\prod_{i=1}^{K}\prod_{l=1}^{M}  h_{j(il)}^{s_{j(il)}},
\end{equation}
where $s_{j(il)}$'s are integers satisfying
\begin{equation}\nonumber
\begin{cases}
s_{j(jl)}=0 & \forall ~ j\in\{1,2,\ldots,K\}~\&~ l\in\{1,\ldots,M\}\\
0\leq s_{j(km)}\leq n-1 &\forall ~ j\in\{1,2,\ldots,K\} ~\& ~j\neq k\\
0\leq s_{j(il)}\leq n & ~ \text{Otherwise.}
\end{cases}
\end{equation}
The set of all transmit directions is denoted by $\mathcal{T}_{km}$.
It is easy to show that the cardinality of this set is
\begin{equation}
L_{km}=n^{K-1}(n+1)^{(KM-1)(K-1)}.
\end{equation}
Clearly, $\mathcal{T}_{km}$ satisfies C1.

We claim that all signals from non-intended cells are aligned at all
base stations. In order to prove the claim, we introduce
$\mathcal{T}_i$ as the set of received direction due to interference
at the $i$'th base stations. Clearly,
$$\mathcal{T}_i=\bigcup_{k=1\& k\neq
i}^{K}\bigcup_{m=1}^{M}(h_{i(km)}\mathcal{T}_{km}).$$

Let us define $\mathcal{T}$ as the set of directions represented by
(\ref{k-user transmit directions1}) and satisfying
\begin{equation}\label{k-user condition1}
\begin{cases}
s_{j(jl)}=0 & \forall ~ j\in\{1,2,\ldots,K\}~\&~ l\in\{1,2,\ldots,M\}\\
0\leq s_{j(il)}\leq n & ~ \text{Otherwise.}
\end{cases}
\end{equation}
We claim that $\mathcal{T}_{i}\subseteq \mathcal{T}$. In fact, all
transmit directions of the $m$'th user in Cell $k$  arrive at
Receiver $i$ multiplied by $h_{i(km)}$. Based on the selection of
transmit directions, however, the maximum power of $h_{i(km)}$ in
all members of $\mathcal{T}_{km}$ is $n-1$. Therefore, none of the
received directions violates the condition (\ref{k-user condition1})
and this proves the claim.

Since $\mathcal{T}$ is not related to the $i$'s base station, one
can conclude that $\mathcal{T}_{i}\subseteq \mathcal{T}$ for all
$i\in\{1,2,\ldots,K\}$. Hence, we deduce that all interfering users
are aligned in the directions of $\mathcal{T}$. Now, $L'_i$ can be
obtained by counting the members of $\mathcal{T}_r$. It is easy to
show that
\begin{equation}
L'_i=(n+1)^{MK(K-1)}.
\end{equation}

The total number of received directions at the $i$'th base stations
is $\sum_{l=1}^M L_{il}+L'_i$. Since $C_1$ and $C_2$ hold at all
base stations, we can obtain the total DOF of the channel as
\begin{IEEEeqnarray}{rl}
r_{\text{sum}}&=\frac{\sum_{k=1}^{K}\sum_{m=1}^{M}L_{km}}{Mn^{K-1}(n+1)^{(KM-1)(K-1)}+(n+1)^{MK(K-1)}+1}\nonumber\\
&=\frac{MKn^{K-1}(n+1)^{(KM-1)(K-1)}}{Mn^{K-1}(n+1)^{(KM-1)(K-1)}+(n+1)^{MK(K-1)}+1}\nonumber\\
&=\frac{MK}{M+\left(\frac{n+1}{n}\right)^{K-1}+\frac{1}{n^{K-1}(n+1)^{(KM-1)(K-1)}}}.
\end{IEEEeqnarray}
Since $n$ can be arbitrary large, we conclude that $\frac{MK}{M+1}$ is achievable for the uplink of a cellular system.

\section{$K\times M$ $X$ Channel}\label{sec X channel}
\subsection{System Model}

\begin{figure}
 \centering
\scalebox{1} 
{
\begin{pspicture}(0,-2.99)(6.4628124,2.97)
\psset{linewidth=0.03cm,arrowsize=0.05291667cm
2.0,arrowlength=1.4,arrowinset=0.4}
\usefont{T1}{ptm}{m}{n}

\def\antenna{%
\begin{pspicture}(1,1)
\pstriangle[gangle=-180.0](0,0.25)(0.6,0.25)
\psline(0,0)(0,-0.60)
\psline(-0.22,-0.60)(0.22,-0.60)
\end{pspicture}
}

\rput[bl](1.7,2.5){\antenna}
\rput[bl](1.7,0.5){\antenna}
\rput[bl](1.7,-2.6){\antenna}

\rput[bl](5.5,1.5){\antenna}
\rput[bl](5.5,-1.5){\antenna}

\psline{->}(.8,2.45)(1.4,2.45)
\rput(0.45,2.45){$x_1$}

\psline{->}(.8,0.45)(1.4,0.45)
\psline{->}(5.7,1.45)(6.3,1.45)
\rput(0.45,0.45){$x_2$}
\rput(6.7,1.5){$y_1$}

\psline{->}(.8,-2.65)(1.4,-2.65)
\psline{->}(5.7,-1.65)(6.3,-1.65)
\rput(0.45,-2.65){$x_K$}
\rput(6.7,-1.65){$y_{K'}$}

\cnode[linecolor=white](2,2.45){.15}{T1}
\cnode[linecolor=white](2,0.45){.15}{T2}
\cnode[linecolor=white](2,-2.65){.15}{Tk}

\cnode[linecolor=white](5.1,1.5){.15}{R2}
\cnode[linecolor=white](5.1,-1.5){.15}{Rk}

\ncline[linecolor=red]{->}{T1}{R2}
\ncput*[nrot=:U,npos=.8]{\tiny $h_{11}$}
\naput*[nrot=:U,npos=.2]{\tiny $\textcolor{red}{m_{11}}$}
\ncline[linecolor=red]{->}{T1}{Rk}
\ncput*[nrot=:U,npos=.8]{\tiny $h_{M1}$}
\nbput*[nrot=:U,npos=.15]{\tiny $\textcolor{red}{m_{M1}}$}
\ncline[linecolor=blue]{->}{T2}{R2}
\ncput*[nrot=:U,npos=.8]{\tiny $h_{12}$}
\naput*[nrot=:U,npos=.2]{\tiny $\textcolor{blue}{m_{12}}$}
\ncline[linecolor=green]{->}{Tk}{Rk}
\ncput*[nrot=:U,npos=.75]{\tiny $h_{MK}$}
\nbput*[nrot=:U,npos=.2]{\tiny $\textcolor{green}{m_{MK}}$}
\ncline[linecolor=green]{->}{Tk}{R2}
\ncput*[nrot=:U,npos=.8]{\tiny $h_{1K}$}
\naput*[nrot=:U,npos=.15]{\tiny $\textcolor{green}{m_{1K}}$}
\ncline[linecolor=blue]{->}{T2}{Rk}
\ncput*[nrot=:U,npos=.7]{\tiny $h_{M2}$}
\nbput*[nrot=:U,npos=.2]{\tiny $\textcolor{blue}{m_{M2}}$}

\psdots[dotsize=0.12](5.5,0.31)
\psdots[dotsize=0.12](5.5,-0.13)
\psdots[dotsize=0.12](5.5,-0.51)
\psdots[dotsize=0.12](1.6209375,-0.75)
\psdots[dotsize=0.12](1.6409374,-1.19)
\psdots[dotsize=0.12](1.6209375,-1.57)
\end{pspicture}
}
\caption{The $K\times M$  $X$ Channel. User $i$ for $i\in
\{1,2,\ldots,K\}$ wishes to transmit an independent message $m_{ji}$
to Receiver $j$ for all $j\in\{1,2,\ldots,M\}$.}\label{fig xchannel}
\end{figure}
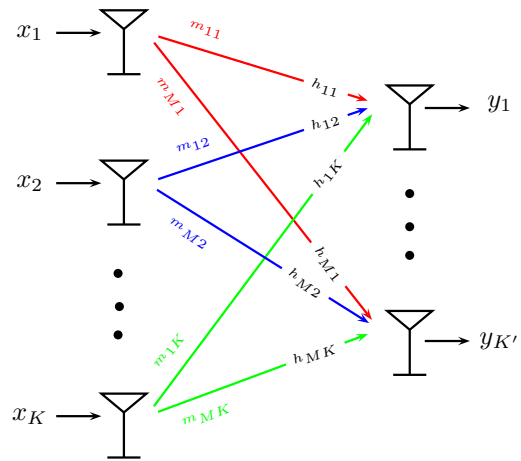

The $K\times M$ $X$ channel models a network in which $K$
transmitters wish to communicate with $M$ receivers. Unlike the
interference channel, each transmitter has a message for each
receiver. In other words, Transmitter $i$ for all
$i\in\{1,2,\ldots,K\}$ wishes to transmit an independent message to
Receiver $j$ for all $j\in\{1,2,\ldots,M\}$. The message transmitted
by Transmitter $i$ and intended for Receiver $j$ is denoted by
$m_{ji}$. The channel's input-output relation can be stated as
follows, see Figure \ref{fig xchannel},
\begin{IEEEeqnarray}{rl}\label{model x}
 y_1 &=h_{11}x_1+h_{12}x_2+\ldots +h_{1K}x_K+z_1,\nonumber\\
 y_2 &=h_{21}x_1+h_{22}x_2+\ldots +h_{2K}x_K+z_2,\nonumber\\
\vdots\ &=\quad  \vdots \quad\qquad \vdots \quad\qquad\ddots\qquad\vdots \\
y_{M} &=h_{M1}x_1+h_{M2}x_2+\ldots +h_{MK}x_K+z_{M},\nonumber
\end{IEEEeqnarray}
where $x_i$ and $y_i$ are input and output symbols of User $i$ for
$i\in\{1,2,\ldots,K\}$, respectively. $z_i$ is Additive White
Gaussian Noise (AWGN) with unit variance for $i\in\{1,2,\ldots,K\}$.
Transmitters are subject to the power constraint $P$. $h_{ji}$
represents the channel gain between Transmitter $i$ and Receiver
$j$. It is assumed that all channel gains are real and time
invariant.

Let $\mathcal{C}_{X}$  denote the capacity region of this channel.
The DOF region associated with the channel can be defined as the
shape of the region in high SNR regimes scaled by $\log \text{SNR}$.
Let us denote the DOF region by $\mathcal{R}_{X}$. We are primarily
interested in the main facet of the DOF region defined as:
\begin{equation}
r_{\text{Xsum}}=\lim_{\text{SNR}\rightarrow\infty}\max_{\mathbf{R}\in
\mathcal{C}_{X}}\frac{\sum_{i=1}^{K}\sum_{j=1}^{M} R_{ij}}{0.5\log\text{SNR}},
\end{equation}
where $R_{ij}$ is an achievable rate for the message $m_{ij}$ and
$\mathbf{R}$ is the set of all achievable rates. The DOF achievable
by the message $m_{ij}$ is denoted by $r_{ij}$.

\subsection{The Total DOF of $\frac{KM}{K+M-1}$ is Achievable}
An upper bound on the DOF of this channel is obtained in
\cite{cadambe2008dfw}. The upper bound states that the total DOF of
the channel is less than $\frac{KM}{K+M-1}$, which means each
message can at most achieve $\frac{1}{K+M-1}$ of DOF. We will show
that this DOF is achievable. To this end, Transmitter $i$ for all
$i\in\{1,2,\ldots,K\}$ transmits $M$ signals along $M$ directions as
follows:
\begin{equation}
x_i=\sum_{j=1}^{M} h_{ji} x_{ji},
\end{equation}
where $x_{ji}$ is the signal carrying the message $m_{ji}$. Let us
focus on the signals intended for Receiver 1, i.e.,
$x_{11},x_{12},\ldots,x_{1K}$. The received signals due to these
transmit signals can be written as
\begin{IEEEeqnarray}{rl}
\tilde{y}_1 &=h_{11}^2x_{11}+h_{12}^2x_{12}+\ldots+h_{1K}^2x_{1K}\nonumber\\
I_{21} &=  (h_{21}h_{11})x_{11}+ (h_{22}h_{12})x_{12}+\ldots+(h_{2K}h_{1K})x_{1K}\nonumber\\
\vdots\ &=\qquad  \vdots \qquad\quad\qquad\qquad \vdots \quad\qquad \ddots \qquad\qquad\vdots \\
I_{M1} &= (h_{M1}h_{11})x_{11}+ (h_{M2}h_{12})x_{12}+\ldots+(h_{M1}h_{1K})x_{1K}.\nonumber
\end{IEEEeqnarray}
Since $x_{11},x_{12},\ldots,x_{1K}$ are not intended for Receiver
$j$ for all $j\in\{2,3,\ldots,M\}$, $I_{j1}$ is a part of
interference at Receiver $j$. We claim that we can align all
interfering signals $x_{11},x_{12},\ldots,x_{1K}$ at all Receivers
$j\in\{2,3,\ldots,M\}$.

Let $\mathbf{H}_1$ denote the set of all coefficients appeared in
$I_{21},I_{31},\ldots,I_{M1}$, i.e.,
$\mathbf{H}_1=\{(h_{21}h_{11}),$ $(h_{22}h_{12}),\ldots,
(h_{M2}h_{12}),h_{M1}h_{1K})\}$. $\mathbf{H}_1$ has $(M-1)K$
members. The set of all monomials with variables in $\mathbf{H}_1$
is denoted by $\mathcal{G}(\mathbf{H}_1)$. Let $\mathcal{T}_1$
denote a subset of $\mathcal{G}(\mathbf{H}_1)$ consisting of
monomials represented by
\begin{equation}
T=\prod_{i=1}^K \prod_{j=1}^{M} (h_{ji}h_{1i})^{s_{ji}},
\end{equation}
where
\begin{equation}\nonumber
\begin{cases}
s_{1i}=0 & \forall ~ i\in\{1,2,\ldots,K\}\\
0\leq s_{ji}\leq n & ~ \text{Otherwise.}
\end{cases}
\end{equation}
Clearly, $\mathcal{T}_1$ has $(n+1)^{(M-1)K}$ members.

The message $m_{1i}$ for $i\in\{1,2,\ldots,K\}$ is transmitted along
directions in $\mathcal{T}_{1i}$ where $\mathcal{T}_{1i}\subset
\mathcal{T}_{1}$. A direction $T$ in $\mathcal{T}_{1i}$ can be
represented as
\begin{equation}
T=\prod_{l=1}^K \prod_{j=1}^{M} (h_{jl}h_{1l})^{s_{jl}},
\end{equation}
where
\begin{equation}\nonumber
\begin{cases}
s_{1l}=0 & \forall ~ l\in\{1,2,\ldots,K\}\\
0\leq s_{ji}\leq n-1 &\forall ~ j\in\{1,2,\ldots,M\} ~\& ~j\neq 1\\
0\leq s_{jl}\leq n & ~ \text{Otherwise.}
\end{cases}
\end{equation}
It is easy to show that the cardinality of $\mathcal{T}_{1i}$ is
$n^{M-1}(n+1)^{(M-1)(K-1)}$. The received directions due to $x_{1i}$
at all receivers belong to  $\mathcal{T}_{1}$. In fact, $x_{1i}$
arrives at receiver $j$ multiplied by $(h_{ji}h_{1i})$ and since the
power of $(h_{ji}h_{1i})$ in all directions in $x_{1i}$ is less than
$n$ we conclude that the received directions are all in
$\mathcal{T}_{1}$. Therefore, all transmit signals are aligned and
the total number of directions in $I_{j1}$ for all
$j\in\{2,3,\ldots,M\}$ is $(n+1)^{(M-1)K}$.

A similar argument can be applied for signals intended for Receiver
$j$ for all $j\in\{2,3,\ldots,M\}$. Therefore, the received signals
can be represented as
\begin{IEEEeqnarray}{rl}\label{received x}
 y_1 &=\tilde{y}_1+I_{12}+I_{13}+\ldots+I_{1M}+z_1,\nonumber\\
 y_2 &=\tilde{y}_2+I_{21}+I_{23}+\ldots+I_{2M}+z_2,\nonumber\\
\vdots\ &=\quad  \vdots \quad\qquad \vdots \quad\qquad\ddots\qquad\vdots \\
y_{M} &=\tilde{y}_M+I_{M1}+I_{M2}+\ldots+I_{(M-1)M}+z_1,\nonumber
\end{IEEEeqnarray}
where $I_{ji}$ is the part of interference caused by all messages
intended for Receiver $i$ at Receiver $j$. Due to symmetry, we only
consider the received directions at Receiver 1. At Receiver 1, there
are $M_1$ interfering signals, each of which consists of at most
$(n+1)^{(M-1)K}$ directions. Therefore, the total number of
interfering directions is $L'_{1}=(M-1)(n+1)^{(M-1)K}$. On the other
hand, $\tilde{y}_1$ consists of $K n^{M-1}(n+1)^{(M-1)(K-1)}$
directions. This is due to the fact that $\tilde{y}_1
=h_{11}^2x_{11}+h_{12}^2x_{12}+\ldots+h_{1K}^2x_{1K}$ and $x_{1i}$
for all $i\in\{1,2,\ldots,K\}$ consists of $
n^{M-1}(n+1)^{(M-1)(K-1)}$ directions. Therefore, the total number
of received directions is $$L=(M-1)(n+1)^{(M-1)K}+K
n^{M-1}(n+1)^{(M-1)(K-1)}.$$ Using Theorem \ref{basic1}, we can
conclude that
\begin{equation}
r_{\text{Xsum}}\geq \frac{\scriptstyle KMn^{M-1}(n+1)^{(M-1)(K-1)}}{\scriptstyle K n^{M-1}(n+1)^{(M-1)(K-1)}+(M-1)(n+1)^{(M-1)K}+1}
\end{equation}
is achievable for the $X$ channel. By rearranging, we obtain
\begin{equation}\label{fit}
r_{\text{Xsum}}\geq \frac{KM}{K+(M-1)\left(\frac{n+1}{n}\right)^{M-1}+\frac{1}{n^{M-1}(n+1)^{(M-1)(K-1)}}}.
\end{equation}
Since (\ref{fit}) holds for all $n$, we obtain
\begin{equation}
r_{\text{Xsum}}=\frac{KM}{K+M-1},
\end{equation}
which is the desired result. In a special case, $M=K$ and the total
DOF is $\frac{K^2}{2K-1}$. This shows that as the number of
transmitter and receivers increases the DOFs of $X$ and GIC behaves
similarly.

\section{Conclusion}\label{sec conclusion}
In this paper, we have considered three static channels, namely the
$K$-user Gaussian Interference Channel (GIC), the uplink channel of
cellular systems, and the $K\times M$ $X$ channel. We have proved
that the total DOF of these systems can be attained by incorporating
real interference alignment in the signaling. We have proved that
single antenna systems can be treated similar to multiple antenna
systems where directions can be used for data transmission and
reception. This result is obtained by proposing a new coding scheme
in which several fractional dimensions are embedded into a single
real line. These fractional dimensions play the role of integral
dimensions in Euclidean spaces. This fact is supported by a recent
extension of Khintchine-Groshev theorem for the non-degenerate
manifolds. The total DOF of the MIMO case as well as the complex
case is also achieved by a simple application of the main result.


\end{document}